\documentclass[a4paper, 10pt]{article}
\usepackage{jheppub}
\usepackage[utf8]{inputenc}
\usepackage[english]{babel}
\usepackage{verbatim}
\usepackage{amsfonts}
\usepackage{amsthm}
\usepackage{mathrsfs}
\usepackage{enumitem}
\usepackage{relsize}
\usepackage{exscale}

\numberwithin{equation}{section}
\DeclareMathOperator{\Tr}{Tr}
\allowdisplaybreaks[4]

\newcommand*\rfrac[2]{{}^{#1}\!/_{#2}}
\notoc 

\newtheoremstyle{mysty}{}{}{}{}{\bfseries}{.}{ }{\thmname{#1}\thmnumber{ #2}\thmnote{ (#3)}}
\theoremstyle{mysty}
\newtheorem{definition}{Definition}[section]

\newtheorem{remark}{Remark}[definition]

\newcommand{\ThreeJ}[6]{\begin{pmatrix}
#1 & #2 & #3 \\
#4 & #5 & #6
\end{pmatrix}}
\newcommand{\SixJ}[6]{\begin{Bmatrix}
#1 & #2 & #3 \\
#4 & #5 & #6 
\end{Bmatrix}}

\begin{document}

\title{Spin foam models and the Duflo Map}
\author[a,c]{Marco Finocchiaro} 
\author[a,b]{Daniele Oriti}
\affiliation[a]{Max Planck Institute for Gravitational Physics (Albert Einstein Institute), Am Muehlenberg 1, D-14476 Potsdam-Golm, Germany, EU}
\affiliation[b]{Arnold-Sommerfeld-Center for Theoretical Physics, Ludwig-Maximilians-Universit\"at, Theresienstrasse 37, D-80333 M\"unchen, Germany, EU}
\affiliation[c]{Institute for Physics, Humboldt-Universit\"at zu Berlin, Newtonstraße 15, 12489 Berlin, Germany, EU}
\emailAdd{marco.finocchiaro@aei.mpg.de}
\emailAdd{daniele.oriti@aei.mpg.de}

\date{\today}
\abstract{We give a general definition of spin foam models, and then of models of 4d quantum gravity based on constraining BF theory. We highlight the construction and quantization ambiguities entering model building, among which the choice of quantization map applied to the $B$ variables carrying metric information after imposing simplicity constraints, and the different strategies for imposing the latter constraints. We then construct a new spin foam model for 4d quantum gravity, using the flux representation 
of states and amplitudes, based on the Duflo quantization map and the associated non-commutative Fourier transform for Lie groups. The advantages of the new model are the geometrically transparent way in which constraints are imposed, and the underlying mathematical properties of the Duflo map itself. Last the presence of a closed analytical formula for the model's amplitudes is another valuable asset for future applications.}

\maketitle

\newpage

\section[Introduction.]{Introduction.}
Spin foam models are a covariant definition of the quantum dynamics of spin network-type structures \cite{Perez:2013uz}. They are thus a covariant counterpart of canonical Loop Quantum Gravity \cite{Rovelli:2004tv, Thiemann:2007wt}, a reformulation of lattice gravity path integrals \cite{Conrady:2008ea, Bonzom:2009hw}, a natural language for state sum models of topological quantum field theories \cite{Turaev:1992hq, Boulatov:1992vp, Ooguri:1992eb} and the perturbative dynamics of group field theories (GFT) \cite{Oriti:2011jm}. In addition to the many results obtained in the context of the mentioned related formalisms (e.g., group field theory and tensor models \cite{Gurau:2012td, Gurau:2012hl}) also directly impacting on spin foam models per se, a lot of work has concentrated on spin foam model building \cite{Barrett:1997gw, Freidel:2007py, Engle:2007wy, Alexandrov:2002br, Kaminski:2009cc, Baratin:2011tx, Baratin:2011hp, Dupuis:2011dh}, and on the semi-classical analysis of the resulting quantum gravity models (for fixed underlying lattice) \cite{Hellmann:2013gva}. More recently, the issues of spin foam renormalization and continuum limit (from both lattice \cite{Bonzom:2013ofa, Dona:2018nev, Delcamp:2016dqo, Dittrich:2016tys, Steinhaus:2018aav, Bahr:2016hwc} and GFT \cite{Carrozza:2013mna, Geloun:2011cy, Geloun:2010vj, Carrozza:2017vkz, Carrozza:2016vsq} perspectives) have become central. Progresses have also been made in generalizing and adapting 
the notions of entanglement entropy and holography, so far mainly in the case of $3d$ quantum gravity models \cite{Delcamp:2016eya, Dittrich:2018xuk, Chirco:2017vhs}. Important steps have also been taken in direction of extracting effective continuum physics out of quantum gravity models, in particular in a cosmological context (both from a canonical perspective \cite{Agullo:2016tjh} and using the GFT reformulation of the same models \cite{Gielen:2014gv, Gielen:2016dss, Oriti:2018qty}). In parallel, we have deepened our understanding of the formal structure of spin foam models, and explored it at a more mathematical level also illuminating the various choices underlying model building. \medskip \\
\noindent This paper tackles the more formal aspects of spin foam construction, as a stepping stone for investigating the more physical ones. First of all, we provide a very general definition of spin foam models and of their construction from their defining building blocks, detailing both the combinatorial aspects and their associated quantum states and amplitudes. The general definition we provide will be a convenient starting point for more model building or the analysis of physical consequences, not relying on any specific representation of quantum states or on model building strategy. Next, we specialize the general construction to spin foam models for 4d quantum gravity, in the Riemannian setting, inspired by the formulation of gravity as a constrained BF theory. We introduce the basic ideas of this construction strategy and provide the general definition of the corresponding models, without focusing exclusively on any one of them, but, on the contrary, highlighting their shared features. We do so in different representations for the quantum states and for the amplitudes. Beside its pedagogical value, the main result of this part is to identify the relevant construction ambiguities and choices characterizing each model as well as the general properties shared by all spin foam models. A good control over both specific and general features will be a useful asset when trying to extract physics from them. Among the shared features, we mention the fact that all spin foam models in this class take the form of non-commutative simplicial gravity path integrals when expressed in the flux (Lie algebra) representation (including the EPRL model which is usually not seen from this perspective). This can greatly facilitate their semi-classical analysis. \cite{Oriti:2014aka} \medskip \\
\noindent Last, we use the previous analysis to construct a {\it new spin foam model} in the same constrained BF class. The model is constructed focusing on the flux/metric representation of spin foam states and amplitudes, and thus relying on the associated tools from non-commutative geometry, notably the non-commutative Fourier transform for Lie groups \cite{Guedes:2013vi, Oriti:2011ac}. A direct advantage of the flux construction is the presence of a closed integral formula for the model's amplitudes in group variables enabling the use of Heat Kernel methods in the renormalizability analysis \cite{Geloun:2011cy}. In this respect, our model parallels the formulation of the one presented in \cite{Baratin:2011hp}. However it employs a different quantization map for the Lie algebra variables of discrete BF theory: the Duflo map. This is another important improvement. On the one hand, the Duflo map has a number of nice mathematical properties (outlined in the Appendix), making it in many ways the natural quantization map for quantum systems based on group-theoretic structures \cite{Duflo:map}. On the other hand, contrary to the quantization map employed in \cite{Baratin:2011hp} (and in some other related quantum gravity literature \cite{Freidel:2005ec}) the Duflo map applies to any semi-simple, locally compact Lie group; thus the model we introduce can be straightforwardly generalised to other dimensions, other model-building strategies and, more immediately relevant, the Lorentzian signature \cite{Oriti:2018bwr}. Last, the Duflo map simplifies the expression of the kernel enforcing the simplicity constraints which, in contrast to \cite{Baratin:2011hp}, allowed us to derive an explicit formula for \textit{fusion coefficients} (\ref{Wi},\,\ref{Wii},\,\ref{FCI}). This is clearly a valuable asset in trying to extract quantitative results about the model and its amplitudes in different regimes by using numerical methods \cite{CFONumerical, Finocchiaro1}.  
%
\section[Abstract formulation of Spin foam models.]{Abstract formulation of Spin foam models.}
In this section we outline the general construction of spin foam models for Riemannian quantum gravity. Spin foam models associate quantum amplitudes to discrete structures, usually in the form of product of amplitudes associated to the lower-dimensional cells of the same lattice. We begin by listing and defining the combinatorial building blocks on which spin foam amplitudes are supported. Then we introduce the Hilbert spaces of boundary states and the set of corresponding amplitudes, as done in \cite{Oriti:2014yla} to which we refer for more details (see also \cite{Kaminski:2010ba, Bahr:2011ey}). This fomulation is also in direct parallel with the GFT completition of spin foam models which implicitly defines their continuum limit suggesting a number of useful field-theoretic tools. \medskip\\
\noindent The discrete building blocks of spin foam models exhibit a \lq molecular\rq\, structure, here introduced in steps.
\begin{definition}[Bisected boundary graph]
A bisected boundary graph $\mathfrak{b}\in\mathfrak{B}$ is an ordered pair $\mathfrak{b} = (\mathcal{V}_{\mathfrak{b}},\mathcal{E}_{\mathfrak{b}})$ forming 
a bipartite graph with vertex partition $\mathcal{V}_{\mathfrak{b}} = \bar{\mathcal{V}}\cup\tilde{\mathcal{V}}$ such that the vertices $\tilde{v}\in\tilde{\mathcal{V}}$ are bivalent. 
\end{definition}
\begin{definition}[Spin foam atom]
A spin foam atom $a\in\mathfrak{A}$ is a triple $\mathfrak{a} = (\mathcal{V}_{\mathrm{a}},\mathcal{E}_{\mathrm{a}},\mathcal{F}_{\mathrm{a}})$, constructed in correspondence with a bisected boundary graph $\mathfrak{b}$, with:
\begin{align}
&\mathcal{V}_{\mathfrak{a}} = \{v\}\cup\mathcal{V}_{\mathrm{b}} \qquad \mathcal{E}_{\mathfrak{a}} = \tilde{\mathcal{E}}\cup\mathcal{E}_{\mathrm{b}} 
\qquad \tilde{\mathcal{E}} = \bigcup_{u}\,\big\{(uv):\,u\in\mathcal{V}_{\mathfrak{b}}\big\} \qquad \mathcal{F}_{\mathfrak{a}} = \bigcup_{\tilde{v}\in\tilde{\mathcal{V}}}\big\{f=(v\bar{v}\tilde{v}):\, (\bar{v}\tilde{v})\in\mathcal{E}_{\mathfrak{b}}\big\}
\end{align}
The set $\mathcal{E}$ contains one edge for each node in $\mathcal{V}_{\mathfrak{b}}$, joining it to the bulk vertex $v$. Last $(v\bar{v}\tilde{v})$ is the boundary face bounded by the edges connecting the three vertices. The set of spin foam atoms is denoted by $\mathfrak{A}$.
\end{definition}
\begin{figure}[ht]
\centering
\includegraphics[scale=0.7]{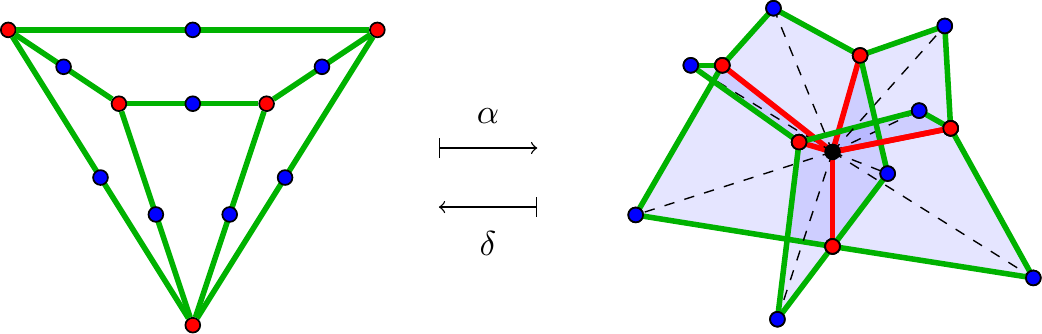}
\caption{A spin foam atom and its bisected boundary graph.}
\label{atom}
\end{figure}
\noindent It is also possible to define two bijective maps: the bulk map $\alpha:\,\mathfrak{B}\rightarrow\mathfrak{A}$ and its inverse, 
the boundary map $\delta:\,\mathfrak{A}\rightarrow\mathfrak{B}$, allowing the set $\mathfrak{A}$ of atoms to be catalogued by the set $\mathfrak{B}$ of bisected boundary graphs \cite{Oriti:2014yla}. 
\begin{definition}[Boundary patch]
A boundary patch $\mathfrak{p}\in\mathfrak{P}$ is a pair $\mathfrak{p}_{\bar{v}} = (\mathcal{V}_{\mathfrak{p}},\mathcal{E}_{\mathfrak{p}})$ such that:
\begin{align}
\mathcal{V}_{\mathfrak{p}} = \{\bar{v}\}\cup\tilde{\mathcal{V}}_{\mathfrak{p}} \qquad \tilde{\mathcal{V}}_{\mathfrak{p}}\neq\emptyset \qquad 
\mathcal{E}_{\mathfrak{p}} = \{(\bar{v}\tilde{v}):\, \tilde{v}\in\tilde{\mathcal{V}}_{\mathfrak{p}}\}
\end{align}
Thus a boundary patch is a graph made by the node $\bar{v}$, all boundary (half)-edges containing it and their end points. Two patches $\mathfrak{p}_{\bar{v}_{1}}$ and $\mathfrak{p}_{\bar{v}_{2}}$ are named {\it bondable} 
if $|\mathcal{V}_{\mathfrak{p}_{1}}| = |\mathcal{V}_{\mathfrak{p}_{2}}|$ and $|\mathcal{E}_{\mathfrak{p}_{1}}| = |\mathcal{E}_{\mathfrak{p}_{2}}|$. Bondable patches can be elementwise identified via a \textit{gluing map} $\gamma:\,\mathfrak{p}_{\bar{v}_{1}}\rightarrow\mathfrak{p}_{\bar{v}_{2}}$.
\end{definition}
%
%
\begin{definition}[Spin foam molecule]
A spin foam molecule $\mathfrak{m}\in\mathfrak{M}$ is a triple of vertices, edges and faces  $\mathfrak{m} = \left(\mathcal{V}_{\mathfrak{m}},\mathcal{E}_{\mathfrak{m}},\mathcal{F}_{\mathfrak{m}}) = (\rfrac{\bigcup_{\mathfrak{a}}\mathcal{V}_{\mathfrak{a}}}{\gamma}, \rfrac{\bigcup_{\mathfrak{a}}\mathcal{E}_{\mathfrak{a}}}{\gamma}, \rfrac{\bigcup_{\mathfrak{a}}\mathcal{F}_{\mathfrak{a}}}{\gamma}\right)$ constructed from a set of spin foam atoms quotiented by a set of gluing maps enforcing the bonding relations between the atoms forming the molecule.
\end{definition}
\begin{figure}[ht] 
\centering
\includegraphics[scale=0.75]{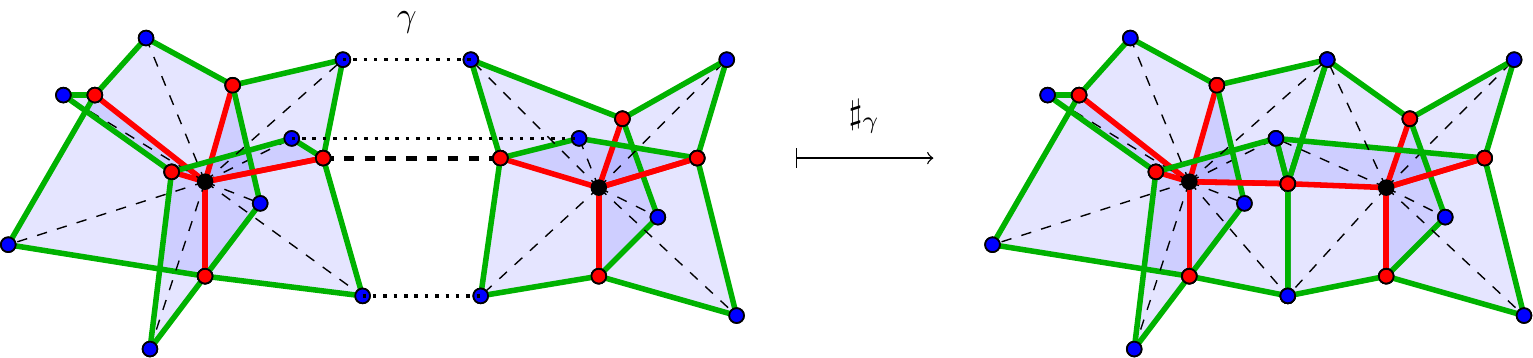}
\caption{The gluing of two atoms along a shared boundary patch to form a molecule.}
\label{molecule}
\end{figure}
%
\begin{definition}[$n$-simplicial structures]
The set of $n$-simplicial molecules $\mathfrak{M}_{\mathrm{S}}$ consists of all molecules obtained as gluings of a single (simplicial) atom $\mathfrak{a}_{\mathrm{S}}$ labelled by the complete graph with $n+1$ vertices $K_{n+1}$.
\end{definition}
\noindent Notice that we call \textit{simplicial}, the above-defined spin foam molecules because each spin foam atom in itself can be canonically understood as the dual 2-skeleton of an $n$-simplex. However, this can be done only locally; it has been proven that not every simplicial spin foam molecule can be associated uniquely to a well-defined simplicial complex, as its dual 2-skeleton \cite{Gurau:2010iu}. While the restriction to simplicial structures is motivated (in addition to simplicity) by the greater geometric understanding of the corresponding models with respect to those based on non-simplicial complexes, we stress that they remain a special case of a more general formalism. The use of arbitrary cellular complexes is suggested by canonical Loop Quantum Gravity \cite{Kaminski:2010ba} and can also be accommodated in the GFT formulation of spin foam models \cite{Oriti:2014yla}, using techniques from dually weighted tensor models.   
\begin{definition}[Spin foam model]
A spin foam model is a quantum theory prescribed by the assignment of a quadruple $(\mathcal{H}_{\mathfrak{p}},\mathfrak{M}_{\mathrm{S}},W,\mathcal{A})$ and defined by a partition function of the following form:
\begin{equation}\label{SF}
\mathcal{Z}_{\mathrm{SF}} = \sum_{\mathfrak{m}\in\mathfrak{M}_{\mathrm{S}}}W(\mathfrak{m})\mathcal{A}(\mathfrak{m})
\end{equation}
Here $\mathcal{H}_{\mathfrak{p}}$ is the Hilbert space associated to each boundary patch of the atoms forming the molecule, $\mathcal{A}(\mathfrak{m})$ is the \textit{spin foam amplitude} assigned to $\mathfrak{m}$ by each given model and $W(\mathfrak{m})$ is a further weight factor in the sum over all molecules. While $\mathcal{A}$ can be motivated, purely by considering the discretization and quantization of some continuum (gravitational) theory the prescription for $W(\mathfrak{m})$ should come from a different line of reasoning. For example, the GFT approach to spin foam models provides a field-theoretic prescription for both of them.
\end{definition}
\noindent The quantum states for which spin foam models define probability amplitudes are associated to the boundary graphs of spin foam molecules. The primary ingredient is the patch Hilbert space, denoted by $\mathcal{H}_{\mathfrak{p}}\equiv\mathcal{H}_{\bar{v}}$. One can then associate an Hilbert space to each spin foam atom $\mathcal{H}_{\mathfrak{a}}$ and to each spin foam molecule $\mathcal{H}_{\mathfrak{m}}$.
\begin{equation}\label{H-patch-tensor}
\mathcal{H}_{\bar{v}} \, =\, \bigotimes_{(\bar{v}\tilde{v})\in\mathcal{E}}\mathcal{H}_{(\bar{v}\tilde{v})} \qquad 
\mathcal{H}_\mathfrak{a}\,=\,\bigotimes_{\mathfrak{p}\in \partial \mathfrak{a}}\mathcal{H}_\mathfrak{p} \qquad
\mathcal{H}_\mathfrak{m}\,=\,\bigotimes_{\mathfrak{p}\in \partial \mathfrak{m}}\mathcal{H}_\mathfrak{p}
\end{equation}
One might also want to define a single Hilbert space for a spin foam model, that would accomadate any possible choice of boundary. This is indeed a crucial issue to tackle the continuum limit and relate the formalism to canonical quantum gravity. From this point of view the simplest proposal is that of a (bosonic) Fock space. This is a natural choice from a QFT/emergent-gravity perspective that sees quantum spacetime as a peculiar quantum many-body system. Another possibility is to define a Hilbert space as the direct sum of all possible graph Hilbert spaces. A third alternative is the one inspired by the canonical LQG construction based on the imposition of \textit{cylindrical equivalence relations}. A comprehensive discussion of these issues can be found in \cite{Oriti:2013aqa}. Let us now turn instead to the construction of the spin foam amplitudes themselves. \medskip\\  
\noindent In order to specify the spinfoam amplitudes $\mathcal{A}_{\mathfrak{m}}$ we need a set of operators defining maps between the various boundary patches' Hilbert spaces. The basic ones are the vertex and glueing operators:
\begin{align}
&\mathsf{O}_{v}\,\,\equiv\,\,\mathsf{V}_{\mathfrak{a}}\,\,\,: \,\,\, \bigotimes_{\mathfrak{p}_{in}\in\partial\mathfrak{a}}\mathcal{H}_{\mathfrak{p}_{in}}\,\,\, \longrightarrow \bigotimes_{\mathfrak{p}_{fin}\in\partial\mathfrak{a}}\mathcal{H}_{\mathfrak{p}_{fin}} \hspace{-20pt}
&\mathscr{V}_{\mathfrak{a}} \,\,\,: \,\,\, \bigotimes_{\mathfrak{p}\in\partial\mathfrak{a}}\mathcal{H}_{\mathfrak{p}}\,\,\, \longrightarrow \,\,\, \mathbb{C} \\
&\mathsf{O}_\gamma\, \equiv\, \mathsf{O}_{e}\,\,:\,\,\mathcal{H}_{\mathfrak{p}_{\bar{v}_{1}}}\longrightarrow\mathcal{H}_{\mathfrak{p}_{\bar{v}_{2}}} \qquad 
&\mathcal{K}_\gamma\, \equiv\, \mathcal{K}_{e} \,\,:\,\, \mathcal{H}_{\mathfrak{p}_{\bar{v}_{1}}}\otimes\mathcal{H}_{\mathfrak{p}_{\bar{v}_{2}}} \longrightarrow\,\,\mathbb{C} \end{align}
The associated functions $\mathscr{V}_{\mathfrak{a}}$ and $\mathscr{K}_{e}$, called the \textit{vertex} and \textit{glueing kernels}, give,  
when applied to any basis in the Hilbert spaces $\mathcal{H}_\mathfrak{p}$, the generalised "matrix elements" of the corresponding operators. The general formula of the spin foam amplitude for a generic molecule, depending on its combinatorial structure, i.e. the connectivity pattern between spin foam atoms and their subcells, is given by:
\begin{equation}\label{SFamplitude}
\mathcal{A}(\mathfrak{m})\, =\, \Tr_{\mathfrak{p}\in\mathfrak{m}}\left(\prod_{e|\mathfrak{m}}\mathcal{K}_{e}\,\prod_\mathfrak{a\in\mathfrak{m}}\mathscr{V}_{\mathfrak{a}}\right)
\end{equation} 
The trace is evaluated over a complete basis in each of the shared patch Hilbert spaces (producing the convolution of the corresponding functions). When the factorized form (\ref{H-patch-tensor}) for the patch Hilbert space is used, the tracing operation take place in each factor associated to each boundary edge of the patch. Following the gluing pattern effected by the gluing maps, one identifies a closed cycle and thus a spin foam face associated to the same patch (for internal patches). Thus the final spin foam amplitude can also be written in terms of individual contributions associated to the faces, edges and vertices of the spin foam molecule. Last these amplitudes, together with an additional combinatorial factor, can be recovered as the perturbative \textit{Feynman amplitudes} of a Group field theory whose propagator and the interaction kernels are given by the same \textit{gluing} and \textit{vertex} kernels of the corresponding (dual) spin foam model \cite{Oriti:2014yla, Oriti:2011jm}.
\section[Spin foam models for constrained BF theory.]{Spin foam models for constrained BF theory.}
Having given the general definitions, let us now focus on the class of gravitational or geometrical Riemannian spin foam models arising from the Holst-Plebanski formulation of General Relativity in $4d$ \cite{Holst:1995pc}. From now on we restric ourselves to simplicial structures. Extensions to the Lorentzian context and to arbitrary cellular complexes can be found in the literature \cite{Perez:2013uz, Oriti:2014yla, Kaminski:2010ba}. In this section, we emphatize and illustrate two points: the construction ambiguities and the universal structure of the resulting amplitudes.
\subsection[Gravity as a constrained BF theory.]{Gravity as a constrained BF theory.}
Our starting point is the Holst-Palatini action without cosmological constant and matter fields, given below:
\begin{equation}
S_{\mathrm{HP}}[e, \omega,\lambda] = \mathlarger{\int}_{\mathcal{M}}\,{}^{*}(e^{I}\wedge e^{J}) \wedge F^{IJ}[\omega] + \frac{1}{\gamma}\,e^{I}\wedge e^{J}\wedge F_{IJ}[\omega]
\label{HP}
\end{equation}
Here $\omega$ is a Spin$(4)$-valued connection one-form field, $F[\omega]$ is the curvature two-form and $e$ is a $\mathfrak{spin}(4)$-valued tetrad one-form field representing an orthonormal frame. The topological term proportional to the Barbero-Immirzi parameter, though irrelevant at the classical level, is vital in the formulation of both Loop Qunatum Gravity (LQG) and Spin foam models. The Holst-Palatini action can be recovered, on the solution of the constraint equations, from a topological BF theory action with additional polynomial constraints $C_{\alpha}[B]$.
\begin{equation}
S[B,\omega,\lambda] = \mathlarger{\int}B^{IJ}\wedge F_{IJ}[\omega] + \lambda^{\alpha}C_{\alpha}[B]
\end{equation}
Several formulation can be given \cite{Gielen:2010cu, DePietri:1998hnx}. In particular the constraints can be taken to be the linear ones:
\begin{equation}
k_{Icdl}(B-\gamma{}^{*}B)^{IJ}_{ab}=0 \qquad k_{Icdl}\equiv \epsilon_{IPKL}e^{P}_{c}e^{K}_{d}e^{L}_{l}
\end{equation}
Ideally one would want to constraint the classical variables first and then quantize the resulting geometric structures. Nevertheless the standard approach is to first quantize the topological discrete BF action and then to impose, directly at the quantum level, a suitable version of the required geometricity constraints hoping to recover the correct gravitational degrees of freedom. On a simplicial lattice $\mathfrak{m}\in\mathfrak{M}_{S}$, the spin connection $\omega$ and the bivector $B$ can be naturally replaced by the holonomy $H_{\bar{v}v}\in\textmd{Spin}(4)$ and the flux variable $X_{\bar{v}\tilde{v}}\in\mathfrak{spin}(4)$ obtained by smearing the original continuum fields on the molecule's subcells of appropriate dimensions.
\begin{align}
&\omega\mapsto H_{\bar{v}v}\equiv\mathcal{P}e^{\int_{e\in\mathcal{E}_{\mathfrak{a}}}\omega_{a}dx^{a}} \qquad B\mapsto X_{\bar{v}\tilde{v}}\equiv\int_{t}B^{IJ}_{ab}dx^{a}\wedge dx^{b} \qquad 
S_{\mathrm{BF}}[H_{\bar{v}v},X_{\bar{v}\tilde{v}}] = \Tr\left(\vec{\zeta}(H_{\bar{v}\bar{v}'})X_{\bar{v}\tilde{v}}\right)  \label{DiscreteBF}
\end{align}
where $H_{\bar{v}\bar{v}'} = H_{\bar{v}v}H_{\bar{v}'v}^{-1}$. The coordinate on the group $\vec{\zeta}(H)$, selected by the choice of the quantization map, dictates the prescription for discretizing the curvature $2$-form \cite{Guedes:2013vi}. Last, by using the canonical decomposition of bivectors into selfdual and anti-selfdual $\mathfrak{su}(2)$ components $X = (x^{-},x^{+})$, the discrete constraints read, 
\begin{equation}\label{LinearSimplicity}
(X^{IJ}_{\bar{v}\tilde{v}} - \gamma {}^{*}X^{IJ}_{\bar{v}\tilde{v}})k_{\bar{v}J} = 0 \qquad k_{\bar{v}}x^{-}_{\bar{v}\tilde{v}}k^{-1}_{\bar{v}} + \beta x^{+}_{\bar{v}\tilde{v}} = 0 \qquad \beta = \frac{\gamma-1}{\gamma + 1}
\end{equation}
where the variable $k_{\bar{v}}$ can be interpreted as the normal to the tetrahedron dual to the patch node $\bar{v}$. The next step is the quantization of the discrete BF theory's phase space and related constraints.
\subsection[The Hilbert space of boundary states.]{The Hilbert space of boundary states.}
The building block of spinfoam boundary states is the patch Hilbert space $\mathcal{H}_{\bar{v}}$.
It admits different realizations as an $L^{2}$ space depending on the choice of variables. Here we consider three distinct choices defining three equivalent formulations of spin foam models, denoted as the \textit{Flux, Holonomy and Spin representations}.
\begin{equation}
\mathcal{H}_{\bar{v}}\equiv L^{2}_{\star}[\mathfrak{spin}(4)^{\times 4}]\otimes L^{2}[S^{3}] \qquad 
\mathcal{H}_{\bar{v}}\equiv L^{2}[\mathrm{Spin}(4)^{\times 4}]\otimes L^{2}[S^{3}] \qquad
\mathcal{H}_{\bar{v}}^{\{J_{i}\}}\equiv \otimes_{i=1}^{4}\mathcal{H}^{J_{i}}\otimes L^{2}[S^{3}]
\label{HilbSp}
\end{equation}
These Hilbert spaces, related by generalized Fourier transforms for Lie groups (i.e. by a change of variables), naturally arise in the quantization of the BF theory's phase space $\mathcal{P}_{\mathrm{BF}} \equiv T^{*}\mathrm{Spin}(4)$. The quantization of the cotangent bundle of a Lie group $T^{*}G$ has been widely studied in the literature \cite{Guedes:2013vi}. Summarizing, the first step is to quantize a maximal subalgebra $\mathscr{A}$ of the Poisson algebra $\mathscr{P}_{G}(C^{\infty}(T^{*}G),\{\bullet,\bullet\},\cdot)$ as an abstract operator ${}^{*}$-algebra $\mathfrak{X}$ by acting on it with a quantization map $\mathcal{Q}:\mathscr{A}\rightarrow\mathfrak{X}$ preserving the commutation relations. The choice of the quantization map (thus the operator ordering) is indeed the \textit{first ambiguity} entering the spin foam model building. Given an abstract quantum algebra of observables $\mathfrak{X}$, the next task is to construct explicit representations $\pi:\mathfrak{X}\rightarrow\mathrm{Aut}(\mathcal{H})$ of it as a concrete operator algebra on suitable Hilbert spaces. The \textit{group} (or \textit{holonomy}) representation $\pi_{G}$ on $\mathcal{H}=L^{2}[G]$ is defined as the one diagonalizing all operators $\mathsf{f}\equiv\mathcal{Q}(f)$ where $f$ are smooth functions on $T^{*}G$. On the other hand the \textit{flux} (or \textit{metric}) representation $\pi_{\mathfrak{g}^{*}}$ on $\mathcal{H} = L^{2}_{\star}[\mathfrak{g}^{*}]$ is the one diagonalizing all flux operators $\mathsf{X}_{i}\equiv\mathcal{Q}(X_{i})$. However since the fluxes do not commute we have to introduce a suitable \textit{star}-product operation which, by deforming the ordinary pointwise multiplication, allows us to satisfy the commutators \cite{Guedes:2013vi}. We can also define the \textit{spin} representation\, $\pi_{J}:\mathfrak{X}\rightarrow\mathcal{B}(\mathcal{H}^{J})$, where $\mathcal{B}(\mathcal{H}^{J})$ denotes the set of bounded linear operators on the group's unitary irreducible representations space $\mathcal{H}^{J}$. \bigskip\\
\noindent In the case of Riemannian spin foam models we chose the structure group $G$ to be the local gauge group of gravity (i.e. $G=\mathrm{Spin}(4)$). By applying the previous quantization procedure (and taking the appropriate tensor product) we immediately recover the single-patch Hilbert spaces (\ref{HilbSp}). From a combinatorial point of view each Hilbert space $\mathcal{H}_{\bar{v}}$ provides the space of one-particle states associated to a single "atom of space", i.e. a quantum tetrahedron in the simplicial setting \cite{Oriti:2017twl, Oriti:2013aqa}. It can be pictured as a fundamental spin-network vertex, represented by a node $\bar{v}\in\bar{\mathcal{V}}$ with $d=4$ 
outgoing half-links and their one-valent end points (i.e. a boundary patch $\mathfrak{p}_{\bar{v}}\in\mathfrak{P}_{\bar{v}}$) labelled, depending on the choice of polarization, by Lie algebra elements, group elements or group representations. The Lie algebra elements can be understood as the covariant smearing of the $B$ $2$-form fields (i.e. fluxes of the B field) across the triangles of the $4$-simplex, the group elements as discretized parallel transports of the BF $1$-form connection along the dual links, and the group representations are quantum numbers labelling eigenstates of the modulus of the $B$ field. Moreover, our quantum (spin network) states are supplemented with an additional variable $k\in S^{3}\simeq \mathrm{Spin}(4)/SU(2)$ interpreted  as the unit 
normal to the tetrahedron in its local $\mathbb{R}^4$ embedding. The presence of such extra variable, and the way it will enter the imposition of the geometricity constraints, implies that our boundary states correspond, more precisely, to {\it projected spin networks} \cite{Dupuis:2010jn}.
%
%
\subsection[Definition and imposition of the geometricity constraints.]{Definition and imposition of the geometricity constraints.}
The geometricity constraints play a crucial role in the formulation of spin foam models for quantum gravity. The closure constraint is a first class constraint, from the canonical point of view. It requires that the sum of the bivectors associated to the four boundary triangles of the same tetrahedron vanishes. This geometrical interpretation is explicit in the flux representation \cite{Baratin:2011tx}. From the group perspective it corresponds to a requirement of covariance under the action of the local gauge group. At the quantum level, it can be implemented by an orthogonal projector $\mathsf{P}_{\mathrm{cl}}$ acting separately on each node of a spin network state, or on each tetrahedron state, taking into account its normal vector $k\in S^{3}$. 
\begin{definition}[Closure constraint operator]
Let $\Psi_{\bar{v}}\equiv\Psi_{k}\in\mathcal{H}_{\bar{v}}$ be a single tetrahedron state and $\mathsf{P}_{\mathrm{cl}}:\ \mathcal{H}_{\bar{v}}\rightarrow\mathcal{H}_{\bar{v}}$ the closure constraint projector. Its expression in the flux representation can be found in \cite{Baratin:2011tx}. In the group representation we have:
\begin{align}
\mathsf{P}_{\mathrm{cl}}(G_{i},\tilde{G}_{i},\tilde{k},k') = \mathlarger{\int}[dH]\,\delta[\tilde{k}^{-1}k']\,\prod_{i=1}^{4}\delta(G^{-1}_{i}H\tilde{G}_{i}) \qquad
\tilde{\Psi}_{k'}(G_{i}) = (\mathsf{P}_{\mathrm{cl}}\Psi_{\tilde{k}})(G_{i}) = \mathlarger{\int}[dH]\Psi_{k'}(HG_{i})
\end{align}
where $k'\equiv H\triangleright k = h^{+}k(h^{-})^{-1}$ denotes the action of a Spin$(4)$ rotation $H$ on the normal $k$. Upon integration over the normal the (extended) closure constraint reduces to the usual gauge invariance of the Ooguri model \cite{Ooguri:1992eb}. The Spin$(4)$ covariance of $\tilde{\Psi}_{k}$ induces its invariance under the action of the stabilizer group SO$_{k}(3)$.
\end{definition}
\noindent The simplicity constraints (\ref{LinearSimplicity}) are second class and cannot be enforced strongly as operatorial equations. Still, in each representation, they correspond to restrictions on the given variables, that are imposed on the spin foam amplitudes via suitable operators. Different strategies have been employed so far leading to the currently known spin foam models. In the following we provide the general form of the constraint operators, highlight the construction ambiguities and indicate briefly the key elements of specific models.
\begin{definition}[Simplicity constraint operator] 
Given a state $\Psi_{\bar{v}}\equiv\Psi_{k}\in\mathcal{H}_{\bar{v}}$, the simplicity constraint operator is defined as a map\;  $\mathsf{S}^{\beta}:\,\mathcal{H}_{\bar{v}}\rightarrow\mathcal{H}_{\bar{v}}$. Its matrix elements in the flux and spin basis are given by:
\begin{align}
&\mathsf{S}^{\beta}_{k}(X_{i},Y_{i}) = \prod_{i=1}^{4}\Big(\delta_{-X_{i}}\star S^{\beta}_{k}\Big)(Y_{i}) \qquad S^{\beta}_{k}(X) = \sum_{J\,j}d_{J}d_{j}\,w(J, j, \beta)\Theta^{J}(X)\star\chi^{j}(kx^{-}k^{-1} + x^{+}) \label{SimplOp}\\
&\mathsf{S}^{J_{1},\dots,J_{4}}_{M_{1},\dots,M_{4}N_{1},\dots,N_{4}}(k, \beta) = \sum_{j_{i}m_{i}}\prod_{i=1}^{4}
C^{j^{-}_{i}j^{+}_{i}j_{i}}_{m^{-}_{i}m^{+}_{i}m_{i}}(\bar{k})
C^{j^{-}_{i}j^{+}_{i}j_{i}}_{n^{-}_{i}n^{+}_{i}m_{i}}(k)w(J_{i},j_{i},\beta) 
\end{align}
The lie algebra characters $\chi^{j}(x)$ and $\Theta^{J}(X)$ are defined in Appendix\,\ref{AppA}.
The function $w(J,j,\beta)$ depends on the chosen imposition criterion. This is the \textit{second main ambiguity} in the construction of spin foam models.
\end{definition}
\begin{remark}[General structure and main properties of $\mathsf{S}^{\beta}$] The simplicity constraint operator $\mathsf{S}^{\beta}$ is not an orthogonal projector for arbitrary values of the Immirzi parameter. Moreover it does not always admit a closed formula for any choice of basis\footnote{It becomes a projector in the limit $\beta\rightarrow\,1$. Remarkably the Barrett-Crane, the Baratin-Oriti and the new model introduced here do admit a closed formula for $\mathsf{S}^{\beta}$ both in the flux, holonomy and spin representations \cite{Baratin:2011tx, Baratin:2011hp}.}. Although different constraint implementation methods yield different definitions of the coefficient $w$, the operator has always the same general structure given above. This feature is a direct consequence of its covariance under the action of Spin$(4)$ and thus of its induced invariance with respect to the stabilizer group SO$_{k}(3)$. In formulas we have:
\begin{equation}
\mathsf{S}^{\beta}(HG_{i}H^{-1}, \tilde{G}_{i}, k) = \mathsf{S}^{\beta}(G_{i}, H\tilde{G}_{i}H^{-1}, k) = \mathsf{S}^{\beta}(G_{i}, \tilde{G}_{i}, k) \qquad H\in\,\mathrm{SO}_{k}(3)
\end{equation}
Due to this underlying rotational symmetry, the operator $\mathsf{S}^{\beta}$ commutes with the closure projector, up to an overall rotation of the normal $k$, thus ensuring a consistent imposition of the geometricity constraints\footnote{Since, in the extended formalism, the closure and simplicity constraint operators commute, we can combine them into a single well defined geometricity operator $\mathsf{G}^{\beta} = \mathsf{P}_{\mathrm{cl}}\circ\mathsf{S}^{\beta}$, enforcing all the required constraints}.
\end{remark}
\noindent Below are some of the current proposals, as found in the literature, to which we refer for more details \cite{Perez:2013uz, Freidel:2007py, Engle:2007wy, Alexandrov:2002br}.
\begin{center}
\small
\begin{tabular}[b]{lll}
\hline\vspace{-5pt} \\
Classes of Models		& $w(J,j,\beta)$ & Method\\
\hline \\
Barrett-Crane,\, $\beta=1$.	& $\delta_{j^{-}j^{+}}\delta_{j0}$.	& Strong imposition. \\
				&	&	\\
EPRL, Alexandrov,\, $\beta<0$.				& $\delta_{j^{-}\,|\beta|j^{+}}\delta_{j\,(1-\beta)j^{+}}$. & Master constraint criterion. \\
Alexandrov,\, $\beta>0$.	& $\delta_{j^{-}\,|\beta|(j^{+}+1)}\delta_{j\,j^{+}(1-|\beta|) - |\beta|}$.	& Master constraint criterion.	\\
				&	&	\\
FK,\, $\beta<0$.			& $\delta_{j^{-}\,|\beta|j^{+}}\delta_{j,\,j^{-}+j^{+}}|C^{j^{-}j^{+}j}_{j^{-}j^{+}j}| = w^{\beta<0}_{\mathrm{EPRL}} $	& Perelomov coherent states. \\
FK,\, $\beta>0$.		& $\delta_{j^{-}\,|\beta|j^{+}}\delta_{j,\,j^{+}-j^{-}}|C^{j^{-}j^{+}j}_{-j^{-}j^{+}j}| $	& Perelomov coherent states.	\\
				&	&	\\
\hline						
\end{tabular}
\end{center}
The single-link fusion coefficient $w$ encodes the simplicity constraints expressed as restrictions on the spins. When no simplicity constraints are imposed, we have $w(J,j,\beta) = 1$ and one recovers the spin foam model for BF theory. It also encodes the choice of the quantization map and thus the operator ordering conventions. Furthermore it can be rescaled by an arbitrary function of the representations $\Delta^{j}_{j^{-}j^{+}}$, compatible with the definition of $\mathsf{S}^{\beta}$ as a map between $L^{2}$ spaces. This is a \textit{third ambiguity} in the construction of spin foam models for $4d$ gravity. The choice of this function is not dictated by the quantization map or by the constraint imposition strategy and thus must be prescribed by hand. Several requirements have been proposed to restrict the allowed choices \cite{Bonzom:2013ofa}.
%
\subsection[Spin foam amplitudes.]{Spin foam amplitudes.}\label{SubSec34}
Once a given prescription to define the quantum geometricity constraints has been selected (yielding a specific expression of the single-link fusion coefficients) we still have to choose how to impose them on the spin foam amplitudes. The main strategies differ according to \textit{where}, i.e. on which ones of the patch Hilbert spaces in the amplitude's definition, the geometricity operator is chosen to act. Once the above decision has been made, we also have to specify whether we intend to act with the geometricity operator itself or with any arbitrary power of the same. This is the \textit{fourth and last ambiguity} in the spin foam construction. When the geometricity operator is an orthogonal projector the above ambiguities are irrelevant. In the following we keep the treatment general and include a parameter indicating such power (which affects the simplicity constraint operator only). Accordingly we can identify three main imposition strategies.
\begin{description}[leftmargin=0pt]
\item[Vertex Hilbert spaces.] Impose the geometricity constraints by acting ($p$ times) with $\mathsf{G}^{\beta}$ 
on the states in the patch Hilbert spaces for each spin foam vertex (atom), before contracting spin foam atoms with one another.
\item[Edge Hilbert spaces.] Enforce the geometricity constraints by acting ($q$ times) with $\mathsf{G}^{\beta}$  on the states 
entering the gluing map, before using it to connect spin foam atoms.
\item[Gluing and Vertex maps.] Implement the geometricity constraints by adopting both the above prescriptions, inserting the geometricity constraint 
both in the vertex and gluing operators.
\end{description}
The above choices lead apriori to different models. Still the amplitudes have the same general structure, given below in the flux and spin bases.
\subsubsection[Simplicial path integral representation of the amplitudes in flux variables.]{Simplicial path integral representation of the amplitudes in flux variables.}
Let us consider a generic (connected) simplicial complex without boundary and the corresponding spin foam molecule $\mathfrak{m}\in\mathfrak{M}_{\mathrm{S}}$.
The spin foam amplitude $\mathcal{A}_{\mathfrak{m}}$ in flux variables is obtained by adopting the corresponding representation of the patch Hilbert spaces and contracting the vertex and gluing kernels according with the general formula ~(\ref{SFamplitude}).
Upon choosing an orientation for each face $f$ the amplitude reads:
\begin{align}
&\mathcal{A}^{\beta}(\mathfrak{m}) = \mathlarger{\int}\bigg[\prod_{v\in\mathcal{V}}\prod_{e\ni v}dH_{ve}\bigg]\bigg[\prod_{e\in\mathcal{E}_{\mathrm{int}}}dk_{e}\bigg]\prod_{f\in\mathcal{F}}\mathcal{A}^{\beta}_{f}(H_{ve}, k_{e}) \\
&\mathcal{A}^{\beta}_{f}(H_{ve},k_{e}) = \mathlarger{\int}\bigg[\frac{d^{6}X_{f}}{(2\pi)^{6}}\bigg]\underset{v,\,e\in f}\bigstar\left(E_{H^{-1}_{ve}}\star S^{\beta\star n}_{k_{e}}\star E_{H_{v'e}}\right)(X_{f})
\end{align}
where $n=2p+2q$ and $S^{\beta\star n}$ denotes the (star-product) nth power of the simplicity function $S^{\beta}$.
We denoted by $H_{ve}$ the parallel transport along an incoming half-edge $e$ and by $X_{f}$ the flux associated to the face $f$. 
%
Notice that the amplitude $\mathcal{A}_{\mathfrak{m}}$ is invariant under the following gauge transformations 
\begin{align}
&H_{ve} \rightarrow \xi_{e}H_{ve}\xi_{v}^{-1} \quad k_{e} \rightarrow \xi^{+}_{e}k_{e}(\xi_{e}^{-})^{-1} \quad X_{f} \rightarrow \xi_{e}X_{f}\xi_{e}^{-1} \qquad \forall\, H,\,\xi\in \textmd{SO}(4),\,\,\, k\in S^3,\,\,\, X_{f}\in\mathfrak{so}(4) 
\end{align}
geometrically interpretated as rotations of all local frames. Such symmetry allows us to drop all the bulk normals from the amplitude, for example by choosing 
the time gauge $\xi_{e} = (k_{e}, \mathbb{I})$. Importantly the general expression of the spinfoam amplitudes in flux variables can be understood as a simplicial path integral for constrained BF theory. In order to see this more clearly we must first commute all simplicity functions $S^{\beta}_{k}$ with the non-commutative plane waves. Hence the amplitude can be rewritten as follows:
\begin{align}
&\mathcal{A}^{\beta}(\mathfrak{m}) = \mathlarger{\int}\bigg[\prod_{f\in\mathcal{F}}\frac{d^{6}X_{f}}{(2\pi)^{6}}\bigg]\bigg[\prod_{e\in\mathcal{E}_{\mathrm{int}}}dk_{e}\bigg]\mathcal{D}_{\beta}^{H_{ve},\,k_{e}}(X_{f})\star\prod_{f\in\mathcal{F}}E_{H_{f}}(X_{f}) \\
&\mathcal{D}^{H_{ve},\,k_{e}}_{\beta}(X_{f}) = \prod_{v\in\mathcal{V}}\prod_{e\ni v}dH_{ve}\prod_{f\in\mathcal{F}}
\bigg[S^{\beta\star n}_{k_{e}}\star\bigg(\underset{e,e'\in f,\,e'\neq e}\bigstar S^{\beta\star n}_{H_{ee'}\triangleright\,k_{e'}}\bigg)\bigg](X_{f})
\end{align}
Here $e$ denotes the reference frame edge while $e'$ any other edge in the same face. The star product runs on all possible couple of edge indeces $(e,e')$ in a given face with the reference edge kept fixed. The simplicity function $S_{H\triangleright k}(\beta)$ imposes on $X_{f}$ the simplicity condition with respect to the rotated normal $H_{ee'}\triangleright k_{e'}$, namely the pull back of the normal $k_{e'}$ to the chosen reference frame. The non-commutative plane waves, collected together for each face of the molecule, gives us the exponential of the discretized BF action. In particular the quantization map dictates the prescription for discretizing the curvature two-form in terms of the holonomy on the dual face, via the choice of coordinates $\zeta(H_{f})$ on the group manifold. For example the FLM map (used in \cite{Baratin:2011hp, Freidel:2005ec, Oriti:2014aka}) discretizes it as the holonomy itself, producing the discrete BF action $\sum_f \Tr\left(X_{f}H_{f}\right)$, while the Duflo map expresses it as the logarithm of the same holonomy (see Appendix\,\ref{AppA}). Last the measure term contains two types of factors.
One, depending only on the fluxes associated to the tetrahedron chosen as reference, impose the simplicity constraints on them, by the chosen prescription characterizing the spin foam model. The remaining part can be viewed as (flux-dependent) 
constraints on the holonomies modifying the Haar measures $dH_{ve}$ on discrete connection parallel transports, used in discrete BF theory. These modifications enforce the requirement that the same discrete connection transports correctly the simplicity constraints across different simplicial frames. The measure also absorbs within it the other construction ambiguities we have mentioned earlier. Its origin can be traced back to the use of the extended states, depending explicitly on the tetrahedral normals and to the consequent generalization of the closure constraint to account for the transformation of the normals (which had been advocated in the spin foam and LQG literature \cite{Alexandrov:2008da, Alexandrov:2011ab}). Both are required in order to ensure a consistent covariant imposition of the geometricity constraints. \medskip\\
\noindent Thus the vacuum amplitudes of any spin foam model (of the class considered) in flux variables take always the form of non-commutative first order simplicial path integrals for a constrained BF theory of the Holst-Plebanski type with a 
discretization of the BF action depending on the chosen quantization map for the fluxes, and a covariant measure  on discrete connection  $\mathcal{D}_{\beta}^{k,\,H,\,X}$ encoding the geometricity constraints, their covariance, and other model-dependent features. We stress once more that this result, first obtained for a specific model in \cite{Baratin:2011tx, Baratin:2011hp} (and for BF theory in \cite{Oriti:2014aka}) is \textit{general}; it is a feature of the flux representation of all models in this constrained BF class, and not the outcome of specific choices in the constraints imposition.
\subsubsection[Spin representation of the amplitudes]{Spin representation of the amplitudes}
The expressions of the amplitudes in the spin basis can be found by taking the non-commutative Fourier transform and the Peter-Weyl decomposition of the above formulas in flux variables. Thus we have:
\begin{align}
\mathcal{A}^{\beta}(\mathfrak{m}) = \sum_{J_{f}j_{ef}}\sum_{I_{ve}i_{e}}\prod_{f\in\mathcal{F}}d_{J_{f}}\prod_{e\in f}d_{j_{ef}}
\prod_{v\in\mathcal{V}}\{15J_{f}\}_{v}\prod_{(\bar{v}v)\equiv e\in\mathcal{E}}d_{I_{ve}}\sqrt{d_{i_{e}}}\,f^{i_{e},p+q}_{I_{ve}}(J_{f},j_{ef},k_{e},\beta)
\end{align}
The fusion coefficients $f^{i}_{I}$ are the matrix elements of a map between Spin$(4)$ and SU$(2)$ intertwiners' spaces.
\begin{align}
&f:\,\mathrm{Inv}_{\mathrm{Spin}(4)}\left[\bigotimes_{l=1}^{4}\mathcal{H}^{j^{-}_{i}}\otimes\mathcal{H}^{j^{+}_{i}}\right]\rightarrow\mathrm{Inv}_{\mathrm{SU}(2)}\left[\bigotimes_{l=1}^{4}\mathcal{H}^{j_{i}}\right] \nonumber \\
&f^{i,p}_{I}(J_{l}, j_{l},k,\beta) = \sum_{M_{l}m_{l}}(\mathcal{I})^{J_{1}J_{2}J_{3}J_{4}I}_{M_{1}M_{2}M_{3}M_{4}}\bigg[\prod_{l=1}^{4}C^{j^{-}_{l}j^{-}_{l}j_{l}}_{m^{-}_{l}m^{-}_{l}m_{l}}(k)w^{p}(J_{l}, j_{l}, \beta)\bigg](\mathcal{I})^{j_{1}j_{2}j_{3}j_{4}i}_{m_{1}m_{2}m_{3}m_{4}} \label{FCI}
\end{align}
Upon using the appropriate recoupling identities the previous definition can be rewritten in terms of NineJ symbols \cite{Varshalovich:1988ye}. The notation goes as follows. We have a Spin$(4)$ representation $J_{f}$ labelling each face, a pair of Spin$(4)$ four-valent intertwiners $I_{ve},\,I_{v'e}$ for every edge and an SO$(3)$ spin $j_{ef}$ for each edge in a given face. Once more, the above formulas are general for this class of constrained BF models, and the specificities of the model lie into the exact form of the single-link fusion coefficients.
\section[Flux variables and the Duflo map: a new spin foam model for quantum gravity.]{Flux variables and the Duflo map: a new spin foam model for quantum gravity.}
We now present a new spin foam model for constrained BF theory, based on the flux/metric procedure introduced in \cite{Baratin:2011tx, Baratin:2011hp} (see also \cite{Guedes:2013vi}) and on the use of the Duflo quantization map (more details in Appendix.\,\ref{AppA}). The choice of the Duflo map is a potentially important aspect since it has a number of attracting mathematical properties. A notable one is that it represents faithfully the subalgebra of invariant functions of the Lie algebra it is applied to, which already makes it the \textit{most natural} quantization map for systems in which the gauge invariance is a key aspect. A second one, possibly even more important in our context, is that it is defined for any semi-simple (locally finite) Lie group, thus in particular it allows an immediate generalization of the construction to the Lorentzian setting, based on the group $SL(2,\mathbb{C})$, which can be found in \cite{Oriti:2018bwr}.
\subsection[Simplicity constraints and non-commutative tetrahedra.]{Simplicity constraints and non-commutative tetrahedra.}
The flux formulation of spin foam models allow us to implement the simplicity condition in a geometrical transparent way: the simplicity constraints are imposed directly on flux variables by \textit{star}-multiplication of states (and the amplitudes) with a non-commutative delta function. Therefore in order to define the model we only need to provide the expressions of the Duflo map and corresponding non-commutative plane waves obtained for SU$(2)$ in \cite{Guedes:2013vi}, which can be trivially extended to Spin$(4)$. The Duflo map is explicitly given by:
\begin{align}
\mathscr{D}:\mathrm{Sym}(\mathfrak{g})\rightarrow U(\mathfrak{g}) \qquad\mathscr{D} = \mathcal{S}\circ\mathcal{J} \qquad \mathcal{J}(X) = \prod_{\pm}\left(\frac{\sinh|x^{\pm}|}{|x^{\pm}|}\right)^{2} \qquad X\in\mathfrak{g}=\mathfrak{spin}(4)
\end{align}
where $\mathcal{S}$ is the symmetric quantization map, $\mathrm{Sym}(\mathfrak{g})$ is the symmetric algebra of $\mathfrak{g}$ and $U(\mathfrak{g})$ denotes its universal enveloping algebra. The plane waves and the integral expansion of the delta function read:
\begin{align}
&E_{G}(X):\mathrm{Spin}(4)\times\mathfrak{spin}(4)\rightarrow\mathbb{C} \qquad E_{G}(X) = E_{g^{-}}(x^{-})E_{g^{+}}(x^{+}) \qquad E_{g^{\pm}}(x^{\pm}) = \frac{|\vec{k}^{\pm}|}{\sin|\vec{k}^{\pm}|}e^{i\vec{k}^{\pm}\cdot\vec{x}^{\pm}} \\
&\delta_{\star}(X-Y) = \mathlarger{\int}dG\,E_{G}(X)E_{G^{-1}}(Y) = \mathlarger{\int}\frac{d^{3}\vec{k}^{-}}{(2\pi)^{3}}\,
\frac{d^{3}\vec{k}^{+}}{(2\pi)^{3}}\,e^{i\vec{k}^{-}\cdot(x^{-}-y^{-})}e^{i\vec{k}^{+}\cdot(x^{+}-y^{+})}
\end{align}
where the vectors $\vec{k}^{\pm}\in\,[0,\pi[$ parametrize the two SU$(2)$ copies of Spin$(4)$ and $\vec{x}^{\pm}\equiv \{x_{i}^{\pm}\}$ are the components of the $\mathfrak{su}(2)$ flux variables $x^{\pm} = x^{\pm}_{i}\sigma^{i}$. Further details can be found in Appendix.\,\ref{AppA}.
\begin{definition}[Simplicity constraint operator] In the flux basis the operator $\mathsf{S}^{\beta}:\mathcal{H}_{\bar{v}}\rightarrow\mathcal{H}_{\bar{v}}$ reads:
\begin{align}
&\mathsf{S}^{\beta}(X_{i}, Y_{i}, k) = \prod_{i=1}^{4}\left(\delta_{-X_{i}}\star S^{\beta}_{k}\right)(Y_{i}) \qquad 
S^{\beta}_{k}(X) = \delta_{-kx^{-}k^{-1}}\left(\beta x^{+}\right) = \mathlarger{\int}du\,E_{k^{-1}uk}(x^{-})E_{u}(\beta x^{+}) \label{SCNew} \\
&\Phi^{\beta}_{k}(X_{i}) = (\mathsf{S}^{\beta}\Psi_{k})(X_{i}) = \mathlarger{\int}\bigg[\prod_{i=1}^{4}\frac{d^{6}Y_{i}}{(2\pi)^{6}}\bigg]\mathsf{S}^{\beta}(X_{i},Y_{i},k)\star\Psi_{k}(Y_{i}) \qquad X,\,Y\in\mathfrak{spin}(4)
\end{align}
Let us now show that the action of this operator is well defined.
\end{definition}
\noindent In order to be able to take the star product with the state $\Psi_{k}$ we need the simplicity functions $S^{\beta}_{k}$ to be in the image of the non-commutative Fourier transform. Therefore we have to rewrite the (\ref{SCNew}) so that the plane waves are evaluated on the same variables $(x^{-}, x^{+})$ labelling the state $\Psi_{k}$. This requirement implies:
\begin{equation}
E_{u}(\beta x) = \Omega(\psi_{u}, \beta)E_{u^{\beta}}(x) \qquad u^{\beta}\in\mathrm{SU}(2) \qquad \beta\in[-1,1]
\label{CC}
\end{equation}
If the previous equality is satisfied for a suitable defined group element $u^{\beta}$ then the simplicity operator is well defined. The above equation admits a unique non-trivial solution given below
\begin{equation}
u_{\beta} = e^{i\frac{\psi_{\beta}}{2}\hat{n}_{\beta}\cdot\vec{\sigma}} \qquad \psi_{\beta} = |\beta|\psi \qquad \hat{n}_{\beta} = \textmd{\textrm{sign}}(\beta)\hat{n} \qquad \Omega(\beta, \psi) = \frac{\sin\frac{|\beta|\psi}{2}}{|\beta|\sin\frac{\psi}{2}}
\end{equation}
where the function $\Omega$ is needed to reconstruct the correct prefactor of the Duflo plane wave. The values of the angles parametrizing the group manifold $S^{3}\simeq\textmd{SU}(2)$, namely $(\psi,\theta,\phi)\in[0,2\pi[\times[0,\pi[\times[0,2\pi[$, ensure the uniqueness of the solution. Thus the simplicity constraint operator becomes:
\begin{equation}
S^{\beta}_{k}(X) = \mathlarger{\int}du\,\Omega(\beta, \psi_{u})E_{k^{-1}uk}(x^{-})E_{u^{\beta}}(x^{+}) = \mathlarger{\int}\mathcal{D}^{\beta}\mathscr{U}\,E_{\mathscr{U}}(X) \qquad \mathcal{D}^{\beta}\mathscr{U} = \Omega(\psi_{u},\beta)du
\label{SCNewII}
\end{equation}
\noindent where, for convenience, we have introduced the following short hand notation $\mathscr{U}=(k^{-1}uk,u^{\beta})$.
\begin{remark}[Properties of $\mathsf{S}^{\beta}_{\mathrm{new}}$]
The simplicity constraint operator $\mathsf{S}^{\beta}$ is not an orthogonal projector for generic values of $\beta$, except for the special cases $\beta=0,1$ (where the simplicity constraints become first class). Due to the properties of the non-commutative plane waves it transforms covariantly under the action of the gauge group Spin$(4)$. Therefore it commutes with the closure projector $\mathsf{P}_{\mathrm{cl}}$, up to a rotation of the normal $k$, and it can be rewritten according to the general formula (\ref{SimplOp}) as a direct calculation would show. Last it does not require any restriction, e.g. any rationality condition, on the Immirzi parameter $\beta$.
\end{remark}
\noindent In the group representation the simplicity operator is given as follows:
\begin{align}
\mathsf{S}^{\beta}(G_{i},\tilde{G}_{i}, k) = \mathlarger{\int}\bigg[\prod_{i=1}^{4}\mathcal{D}^{\beta}\mathscr{U}_{i}\bigg]\prod_{i=1}^{4}\delta\Big[G_{i}\mathscr{U}_{i}\tilde{G}_{i}^{-1}\Big] \label{SElm}
\end{align}
The presence of a closed formula both in the flux and group variables is a peculiar feature of our construction. Upon Peter-Weyl decomposition the simplicity constraint kernel reads:
\begin{align}
&\mathsf{S}^{J_{1},\dots,J_{4}}_{M_{1},\dots,M_{4}N_{1},\dots,N_{4}}(\mathbb{I}, \beta) = \prod_{i=1}^{4}S^{j^{-}_{i}j^{+}_{i}}_{m^{-}_{i}m^{+}_{i}n^{-}_{i}n^{+}_{i}}(\mathbb{I},\beta) = \mathlarger{\int}\bigg[\prod_{i=1}^{4}\mathcal{D}^{\beta}\mathscr{U}_{i}\bigg]\,\prod_{i=1}^{4}D^{j^{-}_{i}}_{m^{-}_{i}n^{-}_{i}}(u_{i})D^{j^{+}_{i}}_{m^{+}_{i}n^{+}_{i}}(u^{\beta}_{i}) \label{Sspin} \\
&S^{j^{-}j^{+}}_{m^{-}m^{+}n^{-}n^{+}}(\beta,\mathbb{I}) = \sum_{jm}C^{j^{-}j^{+}j}_{m^{-}m^{+}m}C^{j^{-}j^{+}j}_{n^{-}n^{+}m}\,w(j^{-}, j^{+}, j, \beta)
\end{align}
where, for shortness, we setted $k=\mathbb{I}$. The single-link fusion coefficient $w(J,j,\beta)$ is given below.
\begin{align}
&w(j^{-}, j^{+}, j, \beta) = \frac{(-1)^{j^{-} + j^{+} + j}}{\pi\,\sqrt{(2j^{-}+1)(2j^{+}+1)}}\,\sum_{a=0}^{\lambda}(\mathrm{Sign}(\beta))^{a}\SixJ{a}{j^{-}}{j^{-}}{j}{j^{+}}{j^{+}}\mathcal{T}_{a}^{j^{-}j^{+}}(\lvert\beta\rvert) \label{Wi} \\
&\mathcal{T}_{a}^{j^{-}j^{+}}(\lvert\beta\rvert) = (-1)^{a}(2a+1)\int_{0}^{2\pi}d\psi\,\Omega(\beta, \psi)\sin^{2}\frac{\psi}{2}\chi_{a}^{j^{-}}(\psi)\chi_{a}^{j^{+}}(\psi_{\beta}) \label{Wii}
\end{align} 
Here $\lambda = 2\,\textmd{Min}(j^{-}, j^{+})$ and $\chi^{j}_{a}$ denotes the generalized character of the SU$(2)$ representations. Further details are provided in Appendicies\,\ref{AppB}-\ref{AppC}. The form factor $w(J,j,\beta)$, characterizing the simplicity constraint imposition in our model, behaves like a modulating weight factor peaking on different configurations, including the ones selected for example by the EPRL model. A detailed numerical investigation of its behaviour and properties in different kinematical regimes will appear in a forthcoming paper \cite{CFONumerical}. The presence of an explicit formula for $w$ is a key asset in trying to extract quantitative consequences of the model's amplitudes, for example the scaling of the leading order radiative corrections of the (connected) $n$-point functions $\mathcal{W}^{(n)}$in the large-j regime. This is the object of a larger project whose results will be published in \cite{Finocchiaro1}. More in general the presence of an explicit closed formula of the model's amplitudes both in the group and in the spin representations is another very interesting property for calculations in the context of spin foam and GFT renormalization, enabling the use of heat kernel and numerical methods in the power counting analysis. In principle this was also a feature of the flux based model \cite{Baratin:2011hp}. In practise however the linear relation between $\psi_{\beta}$ and $\psi$ induced by the Duflo map, contrasted to the non-linear one dictated by the FLM map (namely $\frac{\psi_{\beta}}{2} = \arcsin(|\beta|\sin\frac{\psi}{2})$) simplifies and thus makes possible the evaluation of the integrals. Let us now discuss three interesting limiting cases of these coefficients, thus of the new model.
\begin{description}[leftmargin=0pt]
\item[Barrett-Crane limit.] For $\beta=1$, corresponding to the Plebanski case in which the Holst term disappears, we can immediately evaluate the 
coefficient $w$, to find:
\begin{align}
w(j^{-},j^{+},j,1) = \delta_{j^{-}j^{+}}\delta_{j0}
\end{align}
The simplicity constraint operator also becomes a projector. The result coincides with the restriction on representations defining the Barrett-Crane model in the euclidean signature modulo the choice of the face weights. 
These factors are not determined by the quantization map and thus must be prescribed by hand. In our case, in particular, we recover the  version of the euclidean Barrett-Crane model already presented in \cite{Baratin:2011tx}.
\item[Topological BF theory limit.] For $\beta = 0$, corresponding to $\gamma = 1$, one expects the theory to corresponds to selfdual gravity. The constraint operator $\mathsf{S}^{\beta}$ and the resulting spin foam model are still well defined in this limit despite the fact that the classical constraints have no clear geometrical interpretation. The function $w_{\mathrm{new}}$ is continuous (strictly speaking it has a removable singularity in $\beta=0$, see Appendix.\,\ref{AppC}) and reads:
\begin{align}
w(j^{-},j^{-},j,0) = \frac{2(-1)^{2j^{-}}}{(2j^{-}+1)^{2}}
\end{align}
The simplicity operator acts on the single-tetrahedron state $\Psi_{k}$ by projecting its 
bivectors onto the selfdual part of $\mathfrak{spin}(4)$, and the constrained model reduces to the Ooguri spin foam model for SU$(2)$ BF theory \cite{Ooguri:1992eb}. 
\item[Holst limit.] The case $\beta = -1$, i.e. $\gamma = 0$, corresponds classically to the so-called Holst sector of the Plebanski gravity. This denomination come from the fact that the Holst term of the classical Plebanski-Holst action dominates in this limit. This term vanishes on shell due to the torsion freeness requirement of the spin connection, thus one would be tempted to assume that the corresponding theory is topological. However, the limit in the formal quantum theory for the Plebanski-Holst action in the continuum is subtle, and it has been argued that the resulting quantum theory would correspond to a spin foam 
quantization of 2nd order metric gravity with no torsion \cite{Perez:2013uz, Engle:2007wy, Baratin:2011hp}. In this limit the coefficient $w$ 
is given by:
\begin{align}
w(j^{-},j^{+},j,-1) = \frac{(-1)^{2j^{-}+j}}{2j^{-}+1}\delta_{j^{-}j^{+}}\{j^{-},j^{+},j\}
\end{align}
Therefore for $\beta=-1$ the simplicity constraint operator projects onto simple Spin$(4)$ representation of the type $J = (j^{-},j^{+}) = (l,l)$ without imposing any restriction (apart from the triangular inequalities) on the representation $j$ labelling the decomposition of the tensor product of simple representations $\mathcal{H}^{l}\otimes\mathcal{H}^{l} = \oplus_{j=0}^{2l}\mathcal{H}^{j}$. Moreover it multiplies each pair $(J,j)$ by an extra phase factor $(-1)^{2j^{-}+j}$. Summarizing the resulting spin foam model can be obtained from the ordinary Spin$(4)$ Ooguri model by restricting its representations to simple ones as dictated by the function $w(\beta=-1)$.
\end{description}
\noindent Let us also point out that the simplicity function (\ref{SCNew},\,\ref{SCNewII}) itself (and not just its Fourier modes $S^{J}_{MN}(\beta)$) is well behaved in the cases $\beta=\pm1$ and $\beta=0$ as one can easely check by computing the limits.
\begin{remark}[The role of the quantization map]
In the limit $|\beta|\rightarrow1$ the simplicity function $S^{\beta}_{k}(X)$ reduces, by construction, to an ordinary non-commutative delta function both for the Duflo and for the FLM quantization maps. However the two delta distributions $\delta_{\star_{\mathrm{FLM}}}$ and $\delta_{\star_{\mathscr{D}}}$, seen as functions on the lie algebra, \textit{do not coincide} since they still depend by definition on the star-product, as one can easely check by writing down the corresponding integral expressions in terms of non-commutative plane waves.
\begin{align}
&S^{|\beta|=1}_{\mathrm{new}}(x^{-},x^{+}) = \delta_{\star_{\mathscr{D}}}(x^{-} \pm x^{+}) = \frac{1}{(2\pi)^{2}}\int dS_{2}\int d|\vec{k}||\vec{k}|^{2}e^{i\vec{k}\cdot\vec{x}^{-}}e^{\pm i\vec{k}\cdot\vec{x}^{+}} \\
&S^{|\beta|=1}_{\mathrm{FLM}}(x^{-},x^{+})= \delta_{\star_{\mathrm{FLM}}}(x^{-} \pm x^{+}) = \frac{1}{(2\pi)^{2}}\int dS_{2}\int d|\vec{k}|\sin^{2}|\vec{k}|e^{i\frac{\sin|\vec{k}|}{|\vec{k}|}\vec{k}\cdot\vec{x}^{-}}e^{\pm i\frac{\sin|\vec{k}|}{|\vec{k}|}\vec{k}\cdot\vec{x}^{+}}
\end{align}
Nevertheless they behave as regular Dirac delta functions under integration upon $\star$-multiplication (with respect to the associated $\star$-product) by an arbitrary function $f\in L^{2}_{\star}(\mathfrak{su}(2))$. Therefore, in the above limits $\beta\rightarrow\pm 1$, their Fourier components (\ref{SElm},\,\ref{Sspin} for the Duflo map, \cite{Baratin:2011hp} for the FLM) and thus the spin foam amplitudes for molecules without boundary \textit{will always coincide} regardless of the choice of the quantization map.
The fact that the vacuum spin foam amplitudes coincide is not a puzzling issue. Indeed the physical content of a quantum theory is encoded by the observables rather than by the vacuum amplitudes. As emphatized in \cite{Guedes:2013vi} (see also Appendix.\,\ref{AppA}), there is a precise relation between a quantum operator $\mathsf{O}_{f}$ and the classical function $f_{\star}$ which upon quantization gives $\mathsf{O}_{f}$. Such relation, establishing a connection between the classical phase space and the quantum operators, is controlled by the star-product and thus by the choice of the quantization map. Thus, according to general argument, the quantum obervables have to depend on the operator ordering conventions even in the limit $|\beta|\rightarrow 1$ (though the model's vacuum amplitudes do not). Unlike the previous case, for arbitrary values of $\beta\neq\pm 1$ the model's spinfoam amplitudes do explicitly depend on the operator ordering even for molecules without boundaries. In the flux picture, for istance, the expressions of the simplicity functions and of the vacuum amplitudes for the Duflo map (e.g. the eqs. \ref{SCNewII} and \ref{Afluxnew}) and for the FLM map \cite{Baratin:2011hp} differ by the parametric deformation of the group element $u^{\beta}$ and by the prefactor $\Omega(\psi_{u},\beta)$.  
\end{remark}
%
%
\subsection[The amplitudes of the new model.]{The amplitudes of the new model.}
In this section we present the spin foam amplitudes of the new model. They can be derived straightforwardly from the general definitions introduced in Section.\,\ref{SubSec34}. Since we are going to insert the simplicity constraint operator both in the edge and vertex kernels we introduce, for convenience, the following shorthand notation.
\begin{align}
\mathscr{U}^{(n)} = \bigg(k^{-1}\prod_{a=1}^{n}u_{a}k,\,\prod_{a=1}^{n}u_{a}^{\beta}\bigg) \qquad \mathcal{D}^{\beta}\mathscr{U}^{(n)} = \prod_{a=1}^{n}du_{a}\,\Omega(\beta, \psi_{u_{a}}) 
\qquad u\in \textmd{SU}(2)
\label{notation}
\end{align}
where the integer $n$ counts the number of insertion of the simplicity constraint operator.  
\subsubsection{Flux representation.}
The amplitude $\mathcal{A}_{\mathfrak{m}}$ can be computed exactly as in the general case by taking the appropriate convolution of vertex and gluing kernels. For a closed simplicial complex dual to a molecule $\mathfrak{m}$ without boundary we find:
\begin{align}
&\mathcal{A}^{\beta}_{\mathrm{new}}(\mathfrak{m}) = \mathlarger{\int}\bigg[\prod_{v\in\mathcal{V}}\prod_{e\ni v}dH_{ve}\bigg]\bigg[\prod_{e\in\mathcal{E}_{\mathrm{int}}}dk_{e}\bigg] \prod_{f\in\mathcal{F}_{\mathrm{cl}}}\mathcal{A}^{\beta}_{f}(H_{ve}, k_{e}) \\
&\mathcal{A}^{\beta}_{f}(H_{ve}, k_{e}) = \mathlarger{\int}\bigg[\frac{d^{6}X_{f}}{(2\pi)^{6}}\bigg]\bigg[\prod_{e\in f}\mathcal{D}^{\beta}\mathscr{U}^{(n)}_{ef}\bigg]E_{\big(\underset{e\in f}\prod H^{-1}_{ve}\mathscr{U}^{(n)}_{ef}H_{v'e}\big)}(X_{f}) \label{Afluxnew}
\end{align}
where $n = 2p + 2q$ and $S^{\beta\star n}_{k}$ denotes the star-product $n$th power of the simplicity function $S^{\beta}_{k}$. In deriving the previous formula we relied on the expression (\ref{SCNewII}) and on the properties of the non-commutative plane waves. The amplitude $\mathcal{A}(\mathfrak{m})$ is invariant under simultaneous rotation of all local frames. Therefore it can be evaluated in the time gauge by setting $k_{e} = \mathbb{I}$ for all the internal edges. In order to recast the amplitudes of this model in the form of simplicial path integrals for constrained BF theory, 
we must commute all simplicity functions $S^{\beta}_{k}$ with the plane wave and collect them together. Upon using the Eqs.\,(\ref{EI}) and (\ref{EI}) we find: 
\begin{align}
&\mathcal{A}^{\beta}(\mathfrak{m}) = \mathlarger{\int}\bigg[\prod_{f\in\mathcal{F}_{\mathrm{cl}}}\frac{d^{6}X_{f}}{(2\pi)^{6}}\bigg]\bigg[\prod_{e\in\mathcal{E}_{\mathrm{int}}}dk_{e}\bigg]
\mathcal{D}^{\beta}(H_{ve}, k_{e}, X_{f})\star\prod_{f}E_{H_{f}}(X_{f}) \label{AfluxnewII} \\
&\mathcal{D}^{\beta}(H_{ve}, k_{e}, X_{f}) = \bigg[\prod_{v\in\mathcal{V}}
\prod_{e\ni v}dH_{ve}\bigg]\prod_{f\in\mathcal{F}_{\mathrm{cl}}}\bigg[\underset{e\,e'\in f}\bigstar\,\delta^{\star n}_{(H_{ee'}\triangleright\,k_{e'})x^{-}_{f}(H_{ee'}\triangleright\,k_{e'})^{-1}}\left(\beta x^{+}_{f}\right)\bigg]
\end{align}
where $e$ denotes the reference frame edge while $e'$ any other edge in the same face. The star-product runs on all possible pairs of edge labels $(e,e')$ in a given face with the reference edge kept fixed. The group element $H_{ee'}$ is the parallel transport from the frame $e'$ to the chosen reference frame $e$ and $H_{f}$ is the holonomy of the connection along the face. The non-commutative plane wave in (\ref{AfluxnewII}) gives us the exponential of the discretized BF action and the effective covariant measure satisfies the general properties discussed in Section.\,\ref{SubSec34}. \medskip\\
\noindent Thus we recover again the expression of a non-commutative simplicial path integrals for a constrained BF theory of Holst-Plebansky type, with a discretized BF action depending on the chosen quantization map for the fluxes, and with a measure $\mathcal{D}_{\beta}^{k,\,H}$ capturing the construction choices (and quantization map) characterizing the model. The appearence of an effective measure on the space of discrete connection is a direct consequence of the use of extended states and of the relaxation of the closure constraint (as advocated in the literature \cite{Conrady:2008ea, Dupuis:2010jn, Alexandrov:2008da, Alexandrov:2011ab}) both required to ensure a consistent imposition of the simplicity constraints.
\subsubsection{Group and Spin representation.}
The pure lattice gauge formulation of the model can be found either by non-commutative Fourier transform, exploiting the duality between the metric and the group representations.  For an arbitrary simplicial molecule without boundary the amplitude reads: 
\begin{align}
&\mathcal{A}^{\beta}(\mathfrak{m}) = \mathlarger{\int}\bigg[\prod_{v\in\mathcal{V}}\prod_{e\ni v}dH_{ve}\bigg]\bigg[\prod_{e\in\mathcal{E}_{\mathrm{int}}}dk_{e}\bigg]\prod_{f\in\mathcal{F}_{\mathrm{cl}}}\mathcal{A}^{\beta}_{f}(H_{ve}, k_{e}) \\ 
&\mathcal{A}^{\beta}_{f}(H_{ve}, k_{e}) = \mathlarger{\int}\bigg[\prod_{e\in f}\mathcal{D}^{\beta}\mathscr{U}^{(n)}_{ef}\bigg]\delta\bigg[\prod_{e\in f}H_{ve}\mathscr{U}^{(n)}_{ef}H^{-1}_{v'e}\bigg]
\end{align}
As expected, this is the expression of a lattice gauge theory amplitude for constrained BF theory. The notation goes as follows: the group elements $H_{ve}\in\mathrm{Spin}(4)$ are the parallel transports of the connection along the half-edge $e$ which, upon integration, implement the closure constraint while the variables $U_{ef}$ are the Lagrange multipliers enforcing the linear simplicity constraints. The amplitude is invariant under simultaneous rotation of all local frames and therefore can be evaluated in the time gauge. 
The above amplitude can be equivalently rewritten in the quantum number basis by carrying out all the group integrals.
\newpage
\noindent Upon Peter-Weyl decomposition into irreducible representations it reads: 
\begin{align}
\mathcal{A}^{\beta}(\mathfrak{m}) = \sum_{J_{f}j_{ef}}\sum_{I_{ve}i_{e}}\prod_{f\in\mathcal{F}}d_{J_{f}}\prod_{e\in f}d_{j_{ef}}
\prod_{v\in\mathcal{V}}\{15J_{f}\}_{v}\prod_{(\bar{v}v)\equiv e\in\mathcal{E}}d_{I_{ve}}\sqrt{d_{i_{e}}}\,f_{I_{ve}}^{i_{e},p+q}(J_{f},j_{ef},k_{e},\beta)
\end{align}
As before, the pedices $v,\,e,\,f$ denote the vertices, the edges and the faces of the molecule $\mathfrak{m}$. Uppercase letters denote Spin$(4)$ representations and intertwiners while lowercase letters label SU$(2)$ quantum numbers. Furthermore we have a representation $J_{f}$ for each face, a pair of four-valent intertwiners $I_{ve},\,I_{v'e}$ for every edge and a spin $j_{ef}$ for each edge in a given face. The amplitude is written in terms of the Wigner $15$J symbol \cite{Varshalovich:1988ye, Yutsis:1962vcy} and fusion coefficients (\ref{FCI}) which in turn contain the single-link form factor $w(J,j,\beta)$ in (\ref{Wi},\,\ref{Wii}). Thus they encode the choice of quantization map as well as all the other specificities of our model.
\section[Conclusions.]{Conclusions.}
We have discussed in detail the general structure of spin foam models for quantum gravity, for what concerns both their combinatorial aspects, their quantum states and their quantum amplitudes. We have then specialized to the case of (Riemannian) quantum gravity models in 4d based on the formulation of gravity as a constrained BF theory, which most of the current models belong to \cite{Perez:2013uz}. In the latter case, we have emphasized the construction choices that enter the model building and differentiate different models as well as the mathematical structures common to all of them. We believe that this analysis, beyond its pedagogical usefulness, can be important in providing a clearer base for extracting physics out of these models by focusing both on their specific and on their universal features. Next, we have constructed a new spin foam model, in the same constrained BF class, focusing on the flux/metric representation of spin foam states and amplitudes, and thus relying on the associated tools from non-commutative geometry \cite{Guedes:2013vi}. It is based on the Duflo map \cite{Duflo:map}, for quantizing the Lie algebra variables of discrete BF theory (and thus the metric variables of the gravitational theory). This is important because the Duflo map has a number of nice mathematical properties that make it the natural quantization map for quantum systems based on group-theoretic structures, and because, thanks to them, the model we introduce can be straightforwardly generalise to the Lorentzian signature. We gave the explicit expression of the amplitudes of the new model in flux, group and 
spin representations, and the availability of all of them in closed form is another useful asset, for concrete computations. These concrete computation are the next developments we envisage, based on the new model. First, the model lends itself nicely to numerical evaluations, which allow to compare it straightforwardly with other models in the literature, in particular in the semi-classical (i.e. large-$j$) regime. A first detailed study will be presented in \cite{CFONumerical}. Second, the scaling of the amplitudes, divergences structure and radiative corrections of the new model can be investigated, as a first step towards a more complete renormalization group flow analysis of it. This is also forthcoming work \cite{Finocchiaro1}. The mentioned Lorentzian extension of the new model is another important development that can be already targeted, since we have all the elements for its completion (the mathematical basis for it, i.e. the Lorentzian non-commutative Fourier transform stemming from the Duflo map has been worked out in \cite{Oriti:2018bwr}). Having all these more formal results at hand, we can expect a further strong impulse to the ongoing efforts to extract physics out of spin foam models (also within their GFT reformulation). This is indeed our main goal.
\acknowledgments
The authors thank M.Celoria, L. Sindoni, J. Ben Geloun and G. Chirco for useful discussions and comments.
\newpage
\appendix
\section[The non-commutative Fourier transform and the Duflo map.]{The non-commutative Fourier transform and the Duflo map.} 
\label{AppA}
The quantum geometry underlying both spinfoam models, LQG and GFTs is manifest in the flux representation of their states and amplitudes. This arises from the quantization of the cotangent bundle of a Lie group and relies on the non-commutative Fourier transform. Here we shortly review such topics following the presentation of \cite{Guedes:2013vi} to which we refer for more details. \medskip\\
\noindent The cotangent bundle of a Lie group $T^{*}G$ is a simplectic manifold. Its canonical simplectic structure together with the pointwise multiplication uniquely determines the Poisson algebra and thus the Poisson brackets.
\begin{definition}[Poisson brackets]
Let $\mathscr{P}_{G} = (C^{\infty}(T^{*}G), \{\bullet,\bullet\},\cdot)$ the induced Poisson algebra on the group manifold. The Poisson brackets can be defined as follows:
\begin{align}
\forall f,\,g \in C^{\infty}(T^{*}G) \qquad \{f,g\}\equiv \frac{\partial f}{\partial X_{i}}\mathcal{L}_{i}g - \mathcal{L}_{i}f\frac{\partial g}{\partial X_{i}} + c^{k}_{ij}\frac{\partial f}{\partial X_{i}}\frac{\partial g}{\partial X_{j}}X_{k}
\end{align}
where $\mathcal{L}_{i}$ are the Lie derivative with respect to a basis of right-invariant vector fields, $X_{i}$ are euclidean 
coordinates on the Lie algebra $\mathfrak{g}$, $c^{k}_{ij}$ are the structure constants and repeated lower indeces are summed over. 
\end{definition}
\noindent We now seek to quantize a maximal subalebra $\mathscr{A}$ of this Poisson algebra as an abstract operator $^{*}$-algebra $\mathfrak{X}$.
\begin{definition}[Quantization map]
A quantization map $\mathcal{Q}$ is a linear map between algebras defined as:
\begin{align}
&\mathcal{Q}:\, \mathscr{A}\rightarrow\mathfrak{X} \qquad \forall f \in\mathscr{A}_{G}\subset\mathscr{A}\subset C^{\infty}(G\times\mathfrak{g}^{*}) \qquad\quad\, \mathsf{f}\equiv\mathcal{Q}(f) \quad \mathsf{X}_{i}\equiv\mathcal{Q}(X_{i}) \\
&[\mathsf{f},\mathsf{g}] = 0 \quad [\mathsf{X}_{j}, \mathsf{f}] = \widehat{i\mathcal{L}_{j}f}\in\mathfrak{X}_{G} \quad [\mathsf{X}_{i}, \mathsf{X}_{j}] = ic^{k}_{ij}\mathsf{X}_{k} \qquad \forall\mathsf{f},\,\mathsf{g}\in\mathfrak{X}_{G} = \mathcal{Q}(\mathscr{A}_{G}) \quad \mathsf{X}\in\mathfrak{X}_{\mathfrak{g}^{*}}=\mathcal{Q}(\mathscr{A}_{\mathfrak{g}^{*}})
\end{align}
where $\mathscr{A}_{G}$ is the subalgebra of functions in $\mathscr{A}$ constant on the Lie algebra. Samely $\mathscr{A}_{\mathfrak{g}^{*}}\subset\mathscr{A}$ is the subalgebra of function constant on the group. In general we cannot introduce differentiable coordinates $\zeta^{i}$ on $G$. However they are approximated arbitrarily well by functions in $C^{\infty}(G)$, which allows us to define the operators $\hat{\zeta}^{i}$.
\end{definition}
\noindent Given an abstract quantum algebra of observables $\mathfrak{X}$, the next task is to construct explicit representations $\pi:\mathfrak{X}\rightarrow\mathrm{Aut}(\mathcal{H})$ of it as a concrete operator algebra on suitable Hilbert spaces. The \textit{holonomy} representation $\pi_{G}$ is defined as the one diagonalizing all operators $\mathsf{f}\equiv\mathcal{Q}(f)$ where $f\in C^{\infty}_{c}(G)$. The \textit{flux} representation $\pi_{\mathfrak{g}^{*}}$ is the one diagonalizing all flux operators $\mathsf{X}_{i}\equiv\mathcal{Q}(X_{i})$. Since they do not commute we introduce a suitable \textit{star}-product, deforming the ordinary multiplication, such that the commutation relations are correctly reproduced. 
\begin{definition}[Star product]
A star product, denoted by $\star$, is an operation such that: 
\begin{align}
\forall f_{\star},\,\tilde{f}_{\star}\in\mathscr{A}_{\mathfrak{g}^{*}} \qquad
\mathcal{Q}(f_{\star}\star\tilde{f}_{\star}) = \mathcal{Q}(f_{\star})\mathcal{Q}(\tilde{f}_{\star})
\end{align}
This ensures that $f_{\star}$ can be interpreted as the function that upon quantization gives $f(\mathsf{X}_{i}) = \mathcal{Q}(f_{\star})$, hence establishing a connection between the classical phase space structure and the quantum operators. Therefore the choice of the quantization map determines uniquely the corresponding $\star$-product.
\end{definition}
\begin{definition}[Holonomy and flux representations] Let $\pi_{G}$ and $\pi_{\mathfrak{g^{*}}}$ denote the holonomy and flux representations on the Hilbert spaces $\mathcal{H}=L^{2}(G)$ and $\mathcal{H}=L^{2}_{\star}(\mathfrak{g}^{*})$. The action of the operators $\hat{\zeta}$ and $\mathsf{X}_{i}$ read:
\begin{align}
&\forall \psi\in C_{c}^{\infty}(G)\subset L^{2}(G) \qquad \Big(\pi_{G}(\hat{\zeta}^{i})(\psi)\Big)(g)\equiv \zeta^{i}(g)\psi(g) \qquad \Big(\pi_{G}(\mathsf{X}_{i})(\psi)\Big)(g)\equiv i\mathcal{L}_{i}\psi(g) \\
&\forall\psi\in C^{\infty}_{c}(\mathfrak{g}^{*})\subset L^{2}_{\star}(\mathfrak{g}^{*}) \qquad 
\Big(\pi_{\mathfrak{g}^{*}}(\hat{\zeta}^{i})(\psi)\Big)(x)\equiv -i\frac{\partial}{\partial x_{i}}\psi(x) \qquad \Big(\pi_{\mathfrak{g}^{*}}(\mathsf{X}_{i})(\psi)\Big)(x)\equiv x_{i}\star\psi(X)
\end{align}
The spaces of smooth compactly supported functions on $G$ and $\mathfrak{g}$, dense on the corresponding $L^{2}$ spaces, are closed under the action of the above operators. Moreover the commutators are satisfied upon quantization.
\end{definition}
%
\begin{definition}[Non-commutative Fourier transform]
We assume the existence of a unitary isometric isomorphism $\mathscr{F}:\,L^{2}(G)\rightarrow L^{2}_{\star}(g^{*})$ which 
can be expressed as an integral transform,
\begin{equation}
\tilde{\psi}(x)\equiv \mathscr{F}(\psi)(x)\equiv \int_{G}dg E_{g}(x)\psi(g) \qquad \psi(g) = \mathscr{F}^{-1}(\tilde{\psi})(g)\equiv \int_{\mathfrak{g}^{*}}\frac{d^{D}x}{(2\pi)^{D}}\overline{E_{g}(x)}\star\tilde{\psi}(x)
\end{equation}
where $E_{g}(x)$ denotes the non-commutative plane wave. Their existence has to be verified once an explicit choice of the quantization map has been made. It is a crucial requirements for the existence of $\mathscr{F}$ itself. 
\end{definition}
\noindent Below we list the main properties of the non-commutative plane waves.
\begin{align}
&E_{g}(x+y) = E_{g}(x)E_{g}(y) \qquad E_{gh}(x) = \left(E_{g}\star E_{h}\right)(x) \label{EI}\\
&\overline{E_{g}(x)} = E_{g}(-x) = E_{g^{-1}}(-x) \qquad E_{g}(x)\star E_{h}(x) = E_{h}(x)\star E_{h^{-1}gh}(x) \label{EII}
\end{align}
Last the Lie algebra characters, used in this paper, are defined as follows.
\begin{align}
&\chi^{j}(x) = \int dg\,\chi^{j}(g)E_{g}(x) \qquad x\in\mathfrak{su}(2),\,\,g\in\mathrm{SU}(2) \\
&\Theta^{J}(X) \equiv \chi^{j^{-}}(x^{-})\chi^{j^{+}}(x^{+}) \quad \chi^{j}_{k}(X) \equiv \chi^{j}(kx^{-}k^{-1} + x^{+}) \quad X\equiv(x^{-},x^{+})\in\mathfrak{spin}(4), \,\,\,k\in\mathrm{SU}(2). 
\label{Chi1}
\end{align}
We now outline the basic definitions and properties of the Duflo quantization map and plane wave. 
\begin{definition}[Symmetric quantization map] 
Let $G$ be a semisimple Lie group and $\mathfrak{g}^{*}$ the dual of its Lie algebra. The Symmetric quantization map can be defined as follows:
\begin{align}
\mathcal{S}: \mathrm{Sym}(\mathfrak{g})\longrightarrow U(\mathfrak{g}) \qquad \mathcal{S}(x_{i_{1}},\dots,x_{i_{n}}) = \frac{1}{n!}\sum_{\sigma\in\sigma_{k}}\mathsf{X}_{i_{\sigma_{1}}}\cdots\mathsf{X}_{i_{\sigma_{n}}}
\end{align}
where $\sigma_{k}$ is the symmetric group of order $k$. Here $\mathrm{Sym}(\mathfrak{g})$ is the symmetric algebra of $\mathfrak{g}$ and $U(\mathfrak{g})$ is its Universal enveloping algebra. The symmetric map is not an algebra isomorphism unless $\mathfrak{g}$ is abelian. 
\end{definition}
\begin{definition}[Duflo quantization map] The Duflo map \cite{Guedes:2013vi, Duflo:map} provides an algebra isomorphism between the subalgebra of invariant polynomials under the adjoint group action, denoted by $\mathrm{Sym}(\mathfrak{g})^{\mathfrak{g}}$, and the center of the universal enveloping algebra $U(\mathfrak{g})^{\mathfrak{g}}$. It is defined as follows:
\begin{align}
\mathcal{S}: \mathrm{Sym}(\mathfrak{g})\longrightarrow U(\mathfrak{g}) \qquad 
\mathscr{D} = \mathcal{S}\circ\mathcal{J}^{\frac{1}{2}}(\partial) \qquad \mathcal{J}(x) = \mathrm{det}\left(\frac{\sinh\frac{1}{2}\mathrm{ad}_{x}}{\frac{1}{2}\mathrm{ad}_{x}}\right) = \left(\frac{\sinh|x|}{|x|}\right)^{2}
\end{align} 
where the last expression for $\mathcal{J}$ holds for $x\in\mathfrak{su}(2)$. When applied to exponentials the Duflo map gives:
\begin{align}
&E(g, x) \equiv e_{\star}^{i\vec{k}\cdot \vec{x}} = \eta(g)e^{i\vec{\zeta}(g)\cdot\vec{x}} \qquad \vec{\zeta}(g) = -i\ln(g) = \vec{k}(g) \qquad \eta(g) = \frac{|\vec{k}(g)|}{\sin|\vec{k}(g)|} \qquad  |\vec{k}|\in\,[0,\pi[ \label{PlDII}
\end{align}
that is the expression of the Duflo non-commutative planes wave in the $k$-parametrization.
\end{definition}
\section[Basics of SU(2) harmonic analysis.]{Basics of SU(2) harmonic analysis.} \label{AppB}
In this appendix we collect definitions and identities involving Wigner-$D$ functions, invariant tensors and recoupling coefficients used throughtout the paper and in Appendix.\,\ref{AppC}. Further details can be found in \cite{Varshalovich:1988ye}.
\begin{description}[leftmargin=0pt]
\item[Spherical Harmonics.] The spherical harmonics are a basis in $L^{2}(S^{2})$. Some identities are given below. 
\begin{align}
&Y^{*}_{lm}(\theta, \phi) = Y_{lm}(\theta, -\phi) = (-1)^{m}Y_{l-m}(\theta, \phi) \qquad 
Y_{lm}(\pi-\theta, \pi+\phi) = (-1)^{l}\,Y_{lm}(\theta, \phi) \label{tt}\\
&\mathlarger{\int}_{0}^{\pi}\mathlarger{\int}_{0}^{2\pi}d\theta d\phi\sin\theta Y^{*}_{l'm'}(\theta, \phi)Y_{lm}(\theta, \phi) = \delta_{ll'}\delta_{mm'} \qquad 
\mathlarger{\int}_{0}^{\pi}\mathlarger{\int}_{0}^{2\pi}d\theta d\phi\sin\theta Y_{lm}(\theta, \phi) = \sqrt{4\pi}\delta_{l0}\delta_{m0} \label{tt1}
\end{align}
\item[Wigner matrices.] 
The Wigner matrices are an orthogonal basis in $L^{2}[\mathrm{SU}(2)]$. Their explicit formula reads:
\begin{align}
&D^{j}_{mn}(\psi, \theta, \phi) = \sum_{a=0}^{2j}\sum_{\mu=-a}^{a}(-i)^{a}\left(\frac{4\pi}{2a+1}\right)^{\frac{1}{2}}\left(\frac{2a + 1}{2j + 1}\right)C^{j a j}_{m\mu n}\chi^{j}_{a}(\psi)Y_{a\mu}(\theta, \phi) \label{WMef}
\end{align}
where $\chi^{j}_{a}(\psi)$ denote the generalized character of SU$(2)$ representations introduced below.
\begin{align}
&\chi^{j}_{a}(\psi) = 2^{a}a!\left[\frac{(2j+1)(2j-a)!}{(2j+a+1)}\right]^{\frac{1}{2}}\left[\sin\frac{\psi}{2}\right]^{a}C^{a+1}_{2j-a}\left[\cos\frac{\psi}{2}\right] 
\qquad \chi^{j}_{a}(\psi) = i^{a}\sum_{p=-j}^{j}e^{-ip\psi}C^{jaj}_{p0p} \label{GChef} \\
&\mathlarger{\int}_{0}^{2\pi}d\psi\,\sin^{2}\frac{\psi}{2}\,\chi^{j}_{a}(\psi)\chi^{l}_{a}(\psi) = \pi\,\delta_{jl} \qquad 
\chi^{j}_{a}(0) = \delta_{a0}\chi^{j}(0) \qquad  \chi^{j}_{0}(\psi) = \chi^{j}(\psi) \label{GChgf}
\end{align}
Below we list few useful identities involving the recoupling coefficients.
\item[Recoupling coefficients.] The SU$(2)$ Clebsch-Gordan coefficients satisfy the following identities:
\begin{align}
&\sum_{j_{2}m_{2}}d_{j_{2}}C^{j_{1}j_{2}j_{3}}_{m_{1}m_{2}m_{3}}C^{j_{1}j_{2}j_{3}}_{n_{1}m_{2}n_{3}} = \delta_{m_{1}n_{1}}\delta_{m_{3}n_{3}} \label{CGorthof} \\
&\hspace{-7pt}\sum_{m_{1}m_{2}m_{4}}(-1)^{j_{2}-m_{2}}C^{j_{1}j_{2}j_{3}}_{m_{1}m_{2}m_{3}}C^{j_{4}j_{2}j_{5}}_{m_{4}-m_{2}m_{5}}C^{j_{1}j_{4}j_{6}}_{m_{1}m_{4}m_{6}} = (-1)^{j_{1}+j_{2}+j_{5}+j_{6}}\sqrt{d_{j_{3}}d_{j_{5}}}C^{j_{3}j_{5}j_{6}}_{m_{3}m_{5}m_{6}}\SixJ{j_{2}}{j_{1}}{j_{3}}{j_{6}}{j_{5}}{j_{4}} \label{CGsumf} \\
&C^{j_{1}j_{2}j_{3}}_{m_{1}m_{2}m_{3}} = (-1)^{j_{2}+m_{2}}\left(\frac{2j_{3}+1}{2j_{1}+1}\right)^{\frac{1}{2}}C^{j_{3}j_{2}j_{1}}_{-m_{3}m_{2}-m_{1}} \qquad C^{j_{1}j_{2}j_{3}}_{m_{1}m_{2}m_{3}} = (-1)^{j_{1}+j_{2}+j_{3}}C^{j_{1}j_{2}j_{3}}_{-m_{1}-m_{2}-m_{3}} \label{CGflipf}
\end{align}
Let us now define the SU$(2)$ four-valent intertwiners.
\begin{align}
&\mathlarger{\int}dh\prod_{i=1}^{4}D^{j_{i}}_{m_{i}n_{i}}(h) = \sum_{i=i_{\mathrm{min}}}^{i_{\mathrm{max}}}(\mathcal{I})^{j_{l}i}_{m_{l}}(\mathcal{I})^{j_{l}i}_{n_{l}} \qquad 
(\mathcal{I})^{j_{l}i}_{m_{l}} = \sum_{m}(-1)^{i-m}\sqrt{d_{i}}\ThreeJ{j_{1}}{j_{2}}{i}{m_{1}}{m_{2}}{m}\ThreeJ{i}{j_{4}}{j_{3}}{-m}{m_{4}}{m_{3}}
\end{align}
The Wigner SixJ Symbol, defined in \cite{Varshalovich:1988ye}, satisfies the following two sum rules:
\begin{align}
&\sum_{j_{3}}d_{j_{3}}\SixJ{j_{1}}{j_{2}}{j_{3}}{j_{1}}{j_{2}}{j_{6}} = (-1)^{2j_{6}}\{j_{1},j_{2},j_{6}\} \qquad 
\sum_{j_{6}}(-1)^{j_{1}+j_{2}+j_{6}}d_{j_{6}}\SixJ{j_{1}}{j_{1}}{j_{3}}{j_{2}}{j_{2}}{j_{6}} = \delta_{j_{3}0}\sqrt{d_{j_{1}}d_{j_{2}}} \label{SixJSumRule}
\end{align}
For further details we refer the reader to the standard textbooks \cite{Varshalovich:1988ye, Yutsis:1962vcy}.
\end{description}
\section[The new model's fusion coefficients.]{The new model fusion coefficients.}
\label{AppC}
In this appendix we outline the derivation of the \textit{new model} simplicity functions and fusion coefficients, previously mentioned in the paper. We set $k=\mathbb{I}$ for convenience. 
\begin{description}[leftmargin=0pt]
\item[Simplicity functions.] The matrix elements of the \textit{new model} simplicity constraint operator are given by:
\begin{align}
&S^{j^{-}j^{+}}_{m^{-}m^{+}n^{-}n^{+}}(\beta) = \mathlarger{\int}du\,\Omega(\beta,\psi_{u})D^{j^{-}}_{m^{-}n^{-}}(u)D^{j^{+}}_{m^{+}n^{+}}(u_{\beta}) \qquad \beta\in [-1,1] \qquad \Omega(\beta,\psi) = \frac{\sin\frac{|\beta|\psi}{2}}{|\beta|\sin\frac{\psi}{2}} \label{SimplF} \\
&u_{\beta} = e^{i\frac{\psi_{\beta}}{2}\hat{n}_{\beta}\cdot\vec{\sigma}} \qquad \psi_{\beta} = |\beta|\psi \qquad \hat{n}_{\beta} = \mathfrak{S}(\beta)\hat{n} \qquad \psi,\,\phi\in [0,2\pi]\qquad \theta\in [0,\pi[
\end{align}
where $\mathfrak{S}(\beta)$ is the $\mathrm{sign}$ function. The previous matrix elements can be rewritten as follows: 
\begin{align}
&\mathcal{S}^{j^{-}j^{+}}_{m^{-}m^{+}n^{-}n^{+}}(\beta) = \frac{1}{\pi\,d_{j^{-}}d_{j^{+}}}\,\sum_{a=0}^{\lambda}\sum_{\mu=-a}^{a}
\mathfrak{S}^{a}(\beta)(-1)^{-\mu}\,C^{j^{-}aj^{-}}_{m^{-}\mu n^{-}}C^{j^{+}aj^{+}}_{m^{+}-\mu n^{+}}\mathcal{T}_{a}^{j^{-}j^{+}}(|\beta|) \label{SimplFII}\\
&\mathcal{T}^{j^{-}j^{+}}_{a}(|\beta|) = (-1)^{a}(2a+1)\mathlarger{\int}_{0}^{2\pi}d\psi\,\frac{1}{|\beta|}\sin\frac{\psi}{2}\sin\frac{|\beta|\psi}{2}
\chi^{j^{-}}_{a}(\psi)\chi^{j^{+}}_{a}(|\beta|\psi) \\
&\mathcal{T}_{a}^{j^{-}j^{+}}(|\beta|) = (2a+1)
\sum_{p\,= -j^{-}}^{j^{-}}\sum_{q\,= -j^{+}}^{j^{+}}C^{j^{-}a j^{-}}_{p\, 0\, p}C^{j^{+}a j^{+}}_{q\, 0\, q}\varUpsilon_{pq}(|\beta|) \\
&\varUpsilon_{pq}(|\beta|) = \mathlarger{\int}_{0}^{2\pi}d\psi\,\frac{1}{|\beta|}\sin\frac{\psi}{2}\sin\frac{|\beta|\psi}{2}e^{-i(p + |\beta| q)\psi} \nonumber \\
&= \begin{cases}
\frac{i - ie^{2i\pi|\beta|} + 2\pi|\beta|(|\beta|-1)}{4\beta^{2}(|\beta|-1)}\qquad\quad\,\, \forall\,p,q;\quad 2(p+|\beta|q) = 1-|\beta| \\
\frac{i - ie^{2i\pi|\beta|} - 2\pi|\beta|(|\beta|+1)}{4\beta^{2}(|\beta|+1)}\qquad\quad\,\, \forall\,p,q;\quad 2(p+|\beta|q) = -1-|\beta| \\
-\frac{i - ie^{-2i\pi|\beta|} + 2\pi|\beta|(|\beta|+1)}{4\beta^{2}(|\beta|+1)}\qquad \forall\,p,q;\quad 2(p+|\beta|q) = 1+|\beta| \\
-\frac{i - ie^{-2i\pi|\beta|} - 2\pi|\beta|(|\beta|-1)}{4\beta^{2}(|\beta|-1)}\qquad \forall\,p,q;\quad 2(p+|\beta|q) = -1+|\beta|\\
\frac{8i|\beta|(p+|\beta|q)e^{-2i\pi(p+|\beta|q)}\cos|\beta|\pi - 2\left(-1+\beta^{2}+(p+|\beta|q)^{2}\right)e^{-2i\pi(p+|\beta|q)}\sin|\beta|\pi + 8i|\beta|(p+|\beta|q)}{|\beta|(2|\beta|q + |\beta| + 2p - 1)(2|\beta|q + |\beta| + 2p + 1)(2|\beta|q - |\beta| + 2p - 1)(2|\beta|q - |\beta| + 2p + 1)} \quad\,\mathrm{Otherwise.}
\end{cases}
\end{align}
where $\lambda = 2\,\mathrm{Min}(j^{-},j^{+}) $. To show this we perform the integral in (\ref{SimplF}) by using the Eqs.\.(\ref{WMef}), (\ref{tt}) and (\ref{tt1}). The coefficient $\mathcal{T}^{j^{-}j^{+}}_{a}(|\beta|)$ can be computed from the expressions of the generalized characters (\ref{GChef}). 
\item[Single-link fusion coefficients.] The single link fusion coefficients can be written as follows:
\begin{align} 
&C^{j^{-}j^{+}j}_{n^{-}n^{+}m}\,w(J, j, \beta) \equiv C^{j^{-}j^{+}j}_{m^{-}m^{+}m}\mathcal{S}^{j^{-}j^{+}}_{m^{-}m^{+}n^{-}n^{+}}(\beta) \label{FFO}\\
&w(J, j, \beta) = \frac{(-1)^{j^{-} + j^{+} + j}}{\pi\,\sqrt{(2j^{-}+1)(2j^{+}+1)}}\,\sum_{a}\mathfrak{S}^{a}(\beta)
\SixJ{a}{j^{-}}{j^{-}}{j}{j^{+}}{j^{+}}\mathcal{T}_{a}^{j^{-}j^{+}}(|\beta|) \label{WFO}
\end{align}
To find the expression of $w$ we evaluated the right hand side of the (\ref{FFO}) by using the Eqs.\,(\ref{SimplFII}) and (\ref{CGsumf}).
\item[Limiting cases.] For the values $|\beta| = 1$ and $\beta = 0$ the function $\mathcal{T}^{j^{-}j^{+}}_{a}(|\beta|)$ reads:
\begin{align}
\mathcal{T}^{j^{-}j^{+}}_{a}(\pm 1) = \pi(-1)^{a}(2a+1)\delta_{j^{-}j^{+}} \qquad \mathcal{T}^{j^{-}j^{+}}_{a}(0) = \lim_{\beta\rightarrow 0}\mathcal{T}^{j^{-}j^{+}}_{a}(|\beta|) = \frac{2\pi(-1)^{2j^{-}}(2j^{+}+1)}{(2j^{-}+1)}\delta_{a0}
\label{TLimit}
\end{align}
These identities follows directly from the properties of the generalized characters (\ref{GChgf}). \medskip \\ 
\noindent\textbullet\;\;In the limit $\beta=1$, corresponding to the limit $\gamma\rightarrow\infty$ we recover the Barrett-Crane model:
\begin{align}
S^{j^{-}j^{+}}_{m^{-}m^{+}n^{-}n^{+}}(1) = \frac{(-1)^{m^{-} - n^{-}}}{2j^{-} + 1}\delta_{j^{-}j^{+}}\delta_{m^{-}\,-m^{+}}\delta_{n^{-}\,-n^{+}} \qquad w(j^{-},j^{+},j,1) = \delta_{j^{-}j^{+}}\delta_{j0}
\end{align} 
The first result follows immediately form the definition (\ref{SimplF}) or, in alternative, from the equation (\ref{SimplFII}) upon using the identities (\ref{TLimit}), (\ref{CGorthof}) and (\ref{CGflipf}). The latter one can be found by using the eqs. (\ref{WFO}) and (\ref{SixJSumRule}). \medskip \\
\noindent\textbullet\;\;In the limit $\beta=-1$, we recover the topological Holst model. In particular we find:
\begin{align}
S^{j^{-}j^{+}}_{m^{-}m^{+}n^{-}n^{+}}(-1) = \frac{\delta_{j^{-}j^{+}}}{2j^{-} + 1}\delta_{m^{-}n^{+}}\delta_{n^{-}m^{+}} \qquad w(j^{-},j^{+},j,-1) = \frac{(-1)^{2j^{-} + j}}{(2j^{-}+1)}\delta_{j^{-}j^{+}}\{j^{-},j^{+},j\}
\end{align}
As before the first formula follows from the definition (\ref{SimplF}) or equivalently from the (\ref{SimplFII}) upon using the eqs. (\ref{TLimit}), (\ref{CGorthof}) and (\ref{CGflipf}). The expression of $w$ can be found using again the identities (\ref{WFO}) and (\ref{SixJSumRule}). \medskip \\ 
\noindent\textbullet\;\;In the limit $\beta=0$, corresponding to the SU$(2)$ Ooguri model, the functions $S(\beta)$ and $w(J,j,\beta)$ become:
\begin{align}
&S^{j^{-}j^{+}}_{m^{-}m^{+}n^{-}n^{+}}(0) \equiv  \lim_{\beta\rightarrow\,0^{+}}S^{j^{-}j^{+}}_{m^{-}m^{+}n^{-}n^{+}}(\beta) = \frac{2(-1)^{2j^{-}}}{(2j^{-}+1)^{2}}\delta_{m^{-}n^{-}}\delta_{m^{+}n^{+}} \\
&w(J,j,0) \equiv \lim_{\beta\rightarrow\,0^{\pm}}w(J,j,\beta) = \frac{2(-1)^{2j^{-}}}{(2j^{-}+1)^{2}}
\end{align}
The limits can be easely computed from the definitions (\ref{SimplFII}) and (\ref{WFO}) together with the identity (\ref{TLimit}).
\end{description}

\bibliographystyle{JHEP}
\bibliography{NewModel}

\end{document}